\documentclass[pra,twocolumn,notitlepage,amsmath,superscriptaddress,longbibliography]{revtex4-2}

\usepackage{graphicx}% Include figure files
\usepackage{dcolumn}% Align table columns on the decimal point
\usepackage{bm}% bold math
\usepackage{xcolor}
\usepackage[ruled]{algorithm2e}
\usepackage{amssymb}
\usepackage{amsmath}

\usepackage[colorlinks,
linkcolor=blue,       %%修改此处为你想要的颜色
anchorcolor=blue,  %%修改此处为你想要的颜色
citecolor=blue]{hyperref}

\usepackage{tikz}
\usepackage{pgfplots}
\usepgfplotslibrary{groupplots,units}
\usetikzlibrary{decorations.pathmorphing,spy}

\begin{document}

\title{Atomic clock locking with Bayesian quantum parameter estimation: scheme and experiment} 

\author{Chengyin Han}%
\affiliation{Institute of Quantum Precision Measurement, State Key Laboratory of Radio Frequency Heterogeneous Integration, College of Physics and Optoelectronic Engineering, Shenzhen University, Shenzhen 518060, China}%

\author{Zhu Ma}
\affiliation{Laboratory of Quantum Engineering and Quantum Metrology, School of Physics and Astronomy, Sun Yat-Sen University (Zhuhai Campus), Zhuhai 519082, China}
\affiliation{Institute of Quantum Precision Measurement, State Key Laboratory of Radio Frequency Heterogeneous Integration, College of Physics and Optoelectronic Engineering, Shenzhen University, Shenzhen 518060, China}%
 
\author{Yuxiang Qiu}
\affiliation{Laboratory of Quantum Engineering and Quantum Metrology, School of Physics and Astronomy, Sun Yat-Sen University (Zhuhai Campus), Zhuhai 519082, China}
\affiliation{College of Physics and Electronic Science, Hubei Normal University, Huangshi 435002, China}

\author{Ruihuan Fang}
\author{Jiatao Wu}
\author{Chang Zhan}
\author{Maojie Li}
\affiliation{Laboratory of Quantum Engineering and Quantum Metrology, School of Physics and Astronomy, Sun Yat-Sen University (Zhuhai Campus), Zhuhai 519082, China}
\affiliation{Institute of Quantum Precision Measurement, State Key Laboratory of Radio Frequency Heterogeneous Integration, College of Physics and Optoelectronic Engineering, Shenzhen University, Shenzhen 518060, China}%

\author{Jiahao Huang}
\email{Corresponding author. Email: eqjiahao@gmail.com}
\affiliation{Institute of Quantum Precision Measurement, State Key Laboratory of Radio Frequency Heterogeneous Integration, College of Physics and Optoelectronic Engineering, Shenzhen University, Shenzhen 518060, China}%
\affiliation{Laboratory of Quantum Engineering and Quantum Metrology, School of Physics and Astronomy, Sun Yat-Sen University (Zhuhai Campus), Zhuhai 519082, China}

\author{Bo Lu}%
\email{Corresponding author. Email: lubo1982@szu.edu.cn}
\affiliation{Institute of Quantum Precision Measurement, State Key Laboratory of Radio Frequency Heterogeneous Integration, College of Physics and Optoelectronic Engineering, Shenzhen University, Shenzhen 518060, China}%

\author{Chaohong Lee}
\email{Corresponding author. Email: chleecn@szu.edu.cn}
\affiliation{Institute of Quantum Precision Measurement, State Key Laboratory of Radio Frequency Heterogeneous Integration, College of Physics and Optoelectronic Engineering, Shenzhen University, Shenzhen 518060, China}%
\affiliation{Quantum Science Center of Guangdong-Hong Kong-Macao Greater Bay Area (Guangdong), Shenzhen 518045, China}%

\date{\today}

\begin{abstract}
  
  Atomic clocks are crucial for science and technology, but their sensitivity is often restricted by the standard quantum limit. 
  To surpass this limit, correlations between particles or interrogation times must be leveraged.
  Although the sensitivity can be enhanced to the Heisenberg limit using quantum entanglement, it remains unclear whether the scaling of sensitivity with total interrogation time can achieve the Heisenberg scaling.
  Here, we design an adaptive Bayesian quantum frequency estimation protocol that approaches the Heisenberg scaling and experimentally demonstrate its validity with a cold-atom coherent-population-trapping (CPT) clock.
  In further, we achieve high-precision closed-loop locking of the cold-atom CPT clock by utilizing our Bayesian quantum frequency estimation protocol.
  In comparison to the conventional proportional-integral-differential locking, our Bayesian locking scheme not only yields an improvement of 5.1(4) dB in fractional frequency stability, but also exhibits better robustness against technical noises.
  Our findings not only provide a high-precision approach to lock atomic clocks, but also hold promising applications in various interferometry-based quantum sensors, such as quantum magnetometers and atomic interferometers.
\end{abstract}

\maketitle 
\section{Introduction}
Atomic clocks, which operate by synchronizing a transition between two specific atomic levels with a local oscillator (LO)~\cite{RevModPhys.87.637},
have broad applications in time-keeping, global positioning systems, telecommunications, and also contribute to the search for new physics~\cite{doi:10.1126/science.1192720,Schkolnik_2023,2019Optical,PhysRevLett.125.201302,article_Takamoto}.
Their measurement noises are typically dominated by quantum projection noise (QPN).
The sensitivity of the frequency measurement in atomic clocks utilizing repeated identical Ramsey interferometry is commonly described by $\Delta f_{est}(N,T) \propto \left(N \cdot T \cdot  T_{s}\right)^{-1/2}$. 
Here $N$ is the total particle number, $T$ is the total interrogation time, and $T_s$ is the interrogation time for each Ramsey interferometry. 
This means that the sensitivity obeys the standard quantum limit (SQL) $(\Delta f_{est} \propto N^{-1/2})$ with respect to the total particle number $N$ and the SQL $(\Delta f_{est} \propto T^{-1/2})$ with respect to the total interrogation time $T$.

Utilizing correlations between particles or interrogation times can effectively enhance sensitivity.
Quantum entanglement has been extensively employed to improve sensitivity from SQL to sub-SQL scaling ($\propto N^{-\alpha}$ with $0.5<\alpha\le 1$)~\cite{Robinson2024DirectCO, article_Nichol,Edwin2020,pnas.0901550106,PhysRevLett.117.143004,PhysRevLett.102.033601,Bohnet2014,Esteve2008, Hosten2016MeasurementN1,Pezze2018,Pezze2020,10.1063/5.0204102}.
However, for metrological applications, it is necessary to directly observe stability enhancement without the post-processed removal of technical noise~\cite{Robinson2024DirectCO, Edwin2020,Hosten2016MeasurementN1}.
Alternatively, sensitivity can be improved by utilizing the correlation between measurements with different interrogation times.
Unlike conventional frequentist estimation, the Bayesian estimation relies on updating the current knowledge of parameters after each experiment according to Bayes' law~\cite{Degen2017, Gebhart2023, Lumino2018, PhysRevApplied.16.014035, Cimini2024}. 
The well-designed Bayesian estimation protocols with different interrogation times can improve sensitivity to sub-SQL scaling $(\propto T^{-\alpha})$~\cite{Said2011} or even achieve Heisenberg scaling ($\alpha=1$)~\cite{Wiebe2016,2007Entanglement,Degen2017}.
Besides, Bayesian estimation can efficiently calibrate the experimental parameters and has been demonstrated for quantum computing applications~\cite{Lukens2020,PhysRevResearch.4.013199,PRXQuantum.3.020350,PhysRevApplied.21.014012,Lohani2023}.
The Bayesian quantum parameter estimation has been successfully applied to improve magnetometers~\cite{PhysRevX.9.021019,PhysRevApplied.16.024044,Nusran2012,Bonato2016,Herbschleb2021,Craigie2021, PhysRevA.106.052603}, but \textit{its utilization to enhance atomic clocks has not been reported yet}.

During the closed-loop locking of an atomic clock, the LO frequency is repeatedly referenced to the clock transition frequency of the involved atomic system~\cite{RevModPhys.87.637}.
The closed-loop locking typically involves generating an error signal by comparing the LO frequency with the clock transition frequency, and then applying this error signal to adjust the LO frequency through a servo loop.
Conventionally, the error signal is commonly derived by measuring the bilateral half-maximum points of the frequency-domain Rabi or Ramsey fringes centered around the resonance point. 
The servo loop includes a proportional-integral-differential (PID) controller that works to drive the error signal towards zero, thereby effectively locking the LO frequency to the clock transition frequency.
Typically, the closed-loop locking is achieved with a fixed interrogation time.
Employing Bayesian quantum parameter estimation to integrate the measurements obtained with different interrogation times, one can enhance the sensitivity within a designated interrogation time, enabling the improvement of an atomic clock's stability.
Moreover, Bayesian protocol may inherently offer feedback to dynamically update the LO frequency, making it well-suited for achieving closed-loop locking.
However, \textit{it is still unknown how to utilize Bayesian quantum parameter estimation to achieve the closed-loop locking of an atomic clock}.

In this article, we design and experimentally demonstrate a Bayesian quantum frequency estimation (BQFE) protocol using cold $^{87}$Rb atoms in coherent population trapping (CPT).
We utilize this protocol to achieve high-precision closed-loop locking of a cold-atom CPT clock.
The BQFE protocol is specifically designed to measure the clock frequency of $^{87}$Rb through a sequence of CPT-Ramsey interferometry of exponentially growing interrogation times.
It adaptively adjusts the LO frequency during Bayesian updates.
Our BQFE protocol improves the sensitivity from the SQL scaling $\left(\Delta f_{est} \propto T^{-1/2}\right)$ to the Heisenberg scaling $\left(\Delta f_{est} \propto T^{-1}\right)$.
Furthermore, we successfully lock a cold-atom CPT clock using our BQFE protocol, resulting in a fractional frequency stability of $4.3(2)\times10^{-12}/\sqrt{\tau}$.
This represents an improvement of 5.1(4) dB compared to the fractional frequency stability $1.4(1)\times10^{-11}/\sqrt{\tau}$ obtained by the PID locking.
In addition, our Bayesian locking exhibits better robustness against technical noises.
Beyond enhance the atomic clock locking, our BQFE protocol provides a general promising approach to enhance the robustness and sensitivity of all kinds of interferometry-based quantum sensors. 

This article is organized as follows.
In Sec.~\ref{Sec2}, we introduce the experimental setup and the CPT-Ramsey interferometry for our cold-atom CPT clock system.
In Sec.~\ref{Sec3}, we present the general protocol of Bayesian quantum parameter estimation for measuring the clock frequency and experimentally verify its validity with a cold-atom CPT clock.
In Sec.~\ref{Sec4}, we experimentally demonstrate the high-precision closed-loop locking of a cold-atom CPT clock via our Bayesian quantum frequency estimation protocol.
At last, we give a brief conclusion and discussion in Sec.~\ref{Sec5}.

\section{Cold-atom CPT clock}\label{Sec2}

In this section, we present the experimental setup and  the involved CPT-Ramsey interferometry in our cold-atom CPT clock system.
Instead of using a microwave cavity, CPT~\cite{CPT} interrogates the atomic transitions of alkali atoms optically.
Therefore, CPT atomic clocks have an advantage of low power consumption and small size, which is suitable for developing chip-scale atomic clocks~\cite{10.1063/1.5086319}.
Thermal-atom CPT clocks have demonstrated outstanding short-term stabilities~\cite{PhysRevApplied.7.014018, 10.1063/1.4977955, Yun_2021}, but the stabilities are degraded for averaging times higher than 100 or 1000 s.
This degradation is generally caused by laser intensity and frequency light-shift effects, cell temperature, or pressure effects~\cite{10.1063/1.5030009}.
In comparison to thermal-atom CPT clocks, cold-atom CPT clocks are promising to eliminate the degradation of mid- and long-term stabilities~\cite{PhysRevApplied.8.054001, 10.1063/1.5001179,Liu_2022}.
On one hand, cold atoms stay in a clear environment and thus there are no pressure frequency shifts caused by buffer gases.
On the other hand, cold atoms could be interrogated by longer Ramsey free-evolution time, which can greatly reduce the light frequency shifts that are inversely proportional to the Ramsey free-evolution time.

\begin{figure*}
	\includegraphics[width=1\linewidth]{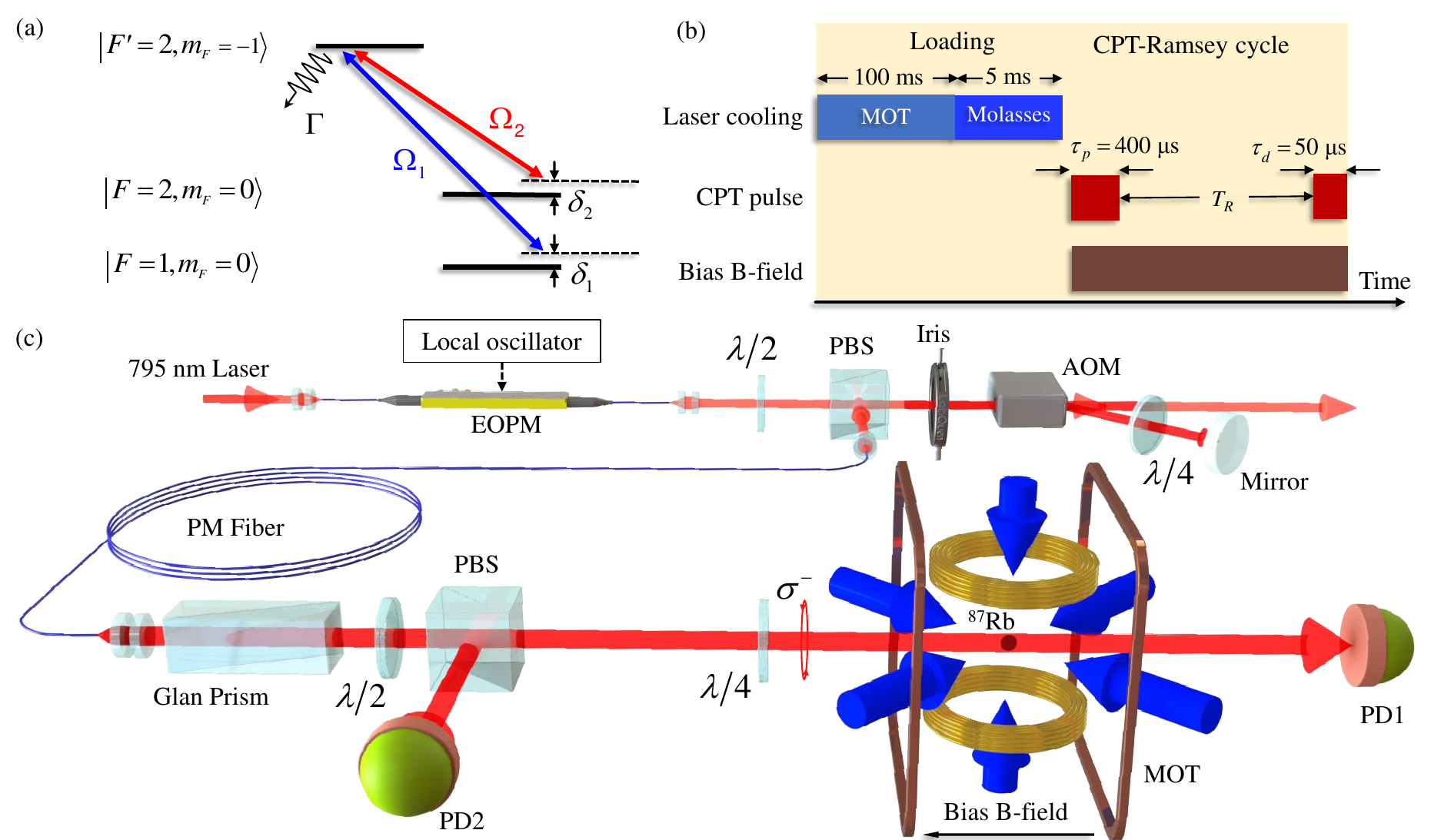}
	\caption{\label{fig_expsetup}
(a) A typical three-level $\Lambda$ system in $^{87}\rm Rb$.
In our experiment, the two clock states $(\vert F=1,m_F=0\rangle\ $ and $ \vert F=2,m_F=0\rangle)$ are coupled through the excited state $\vert F^{\prime}=2,m_F=-1\rangle$.
(b) The timing sequence for an experimental cycle.
The loading process contains 100 ms MOT and 5 ms molasses. The CPT-Ramsey process includes a 400 $\rm \mu s$ preparation pulse $\tau_p$, a Ramsey time $T_R$ and a 50 $\rm \mu s$ detection pulse $\tau_d$.
The bias magnetic field aligned with the CPT laser beam is applied in the CPT-Ramsey process to split the Zeeman sublevels.
(c) Experimental setup.
The CPT beam is generated and modified by a fiber-coupled EOPM, an AOM and a Glan prism.
To increase signal robustness to laser intensity noise, the CPT beam is separated into a transmission beam and a reference beam by a half-wave plate and a PBS.
The $\sigma-\sigma$ configuration of the CPT beam is realized by a quarter-wave plate according to the bias B-field.
AOM: acousto-optic modulator, EOPM: electro-optic phase modulator, PM: polarization maintaining, PBS: polarization beam splitter, PD: photodetector, $\lambda /2$: half-wave plate, $\lambda /4$: quarter-wave plate.}
\end{figure*}

\subsection{Experimental setup}

We utilize a single-$\Lambda$ CPT scheme to avoid time-dependent intervening between multi-path CPT in the lin$\parallel$lin or lin$\perp$lin scheme~\cite{lin11lin_formula_2,lin11lin_realization_1,lin11lin_realization_3,linperlin_realization_1,linperlin_realization_4}.
The single-$\Lambda$ CPT connects the clock states $\vert 1\rangle=\vert F=1,m_F=0\rangle\ \rm{and}\ \vert 2\rangle=\vert F=2, m_F=0\rangle$ through the excited state $\vert 3\rangle=\vert F^{\prime}=2,m_F=-1\rangle$, see Fig. \ref{fig_expsetup}(a).
$\Gamma$ is the decay rate of spontaneous emission from the excited state to the ground states.
Two laser fields with a frequency difference equal to the hyperfine transition frequency $\omega_{hf}$ between $\vert 1\rangle$ and $\vert 2\rangle$ are applied to couple the $\Lambda$ system.
The detuning $\delta_1$ and $\delta_2$ from optical resonances correspond respectively to single photon detuning.
The two-photon detuning $\Delta$ and the average detuning $\delta$ of the laser field are defined as $\Delta=\delta_1-\delta_2$ and $\delta=(\delta_1+\delta_2)/2$.
The Rabi frequency of the transition from the two ground states to the common excited state is expressed as $\Omega_1$ and $\Omega_2$, respectively.
The experimental setup of our cold-atom CPT system is similar to our previous work~\cite{Fang2021,Han2021}, and we give a brief introduction here.

In our experiment, a three-dimensional magneto-optical trap (MOT) is adopted for $100$ ms followed by a $5$ ms molasses, which can trap and cool approximately $10^7$ $^{87}$Rb atoms to $20$ $\rm \mu K$.
At this moment, the atoms are fallen freely and interrogated by left-circularly polarized CPT light in a bias magnetic field of $35$ mG.
The CPT beam is generated from a laser modulated by a fiber-coupled electro-optic phase modulator (EOPM) at $6.835$ GHz, which is equal to the $^{87}$Rb ground-state hyperfine splitting frequency.
The optical carrier and positive first-order sideband are set as equal intensity and couple the clock states to the excited state.
An acousto-optic modulator (AOM) is used to generate the CPT pulses.
After a fiber, a Glan prism is used to purify the polarization.
We use a timing diagram of a $400$ $\rm \mu s$ CPT preparation pulse $\tau_p$ followed by an interrogation time $T_R$ and a $50$ $\rm \mu s$ detection pulse $\tau_d$ for an experimental cycle, see Fig.~\ref{fig_expsetup}~(b).
The CPT beam is equally separated into two beams by a half-wave plate and a polarization beam splitter.
One beam, whose polarization is changed to $\sigma^{-}$ by a quarter-wave plate, transmits through the atoms and is detected by the CPT photodetector as $S_T$, see PD1 in Fig.~\ref{fig_expsetup}~(c).  
While the other beam is detected by the photodetector as $S_N$, see PD2 in Fig.~\ref{fig_expsetup}~(c).
The transmission signals (TSs) are given by $S_{TS}=S_T/S_N$, which can reduce the effect of intensity noise on the CPT-Ramsey signals.

\subsection{CPT-Ramsey interferometry}

Our cold-atom CPT clock operates based upon the CPT-Ramsey interferometry.
The CPT-Ramsey interferometry is similar to the conventional Ramsey interferometry via two-level system and observed CPT-Ramsey interference pattern can reveal the hyperfine resonance in the ground states~\cite{doi:10.1080/23746149.2024.2317896}.
For the D1 line of $^{87}$Rb atom including two clock states $(\vert 1 \rangle = \vert F=1,m_F=0\rangle $ and $ \vert 2 \rangle = \vert F=2,m_F=0\rangle)$ and a common excited state ($\vert 3 \rangle = \vert F^{\prime},m_F=-1\rangle$), this three-level system can be described by the density matrix $\rho_{ij}(i,j=1,2,3)$ obeying the Liouville equation \cite{PhysRevA.84.062502},
\begin{eqnarray}
	 \frac{d}{dt}\rho_{ij}=\frac{1}{i\hbar}\sum_{k}^{}(H_{i,k}\rho_{k,j}-\rho_{i,k}H_{k,j})+\mathcal{R}\rho_{ij}.
	\label{eq:master_equation}
\end{eqnarray}
The coupling of these three states with two coherent radiation fields is described by the following Hamiltonian under the rotating wave approximation (RWA),
\begin{eqnarray}
	H=
	\left(
	\begin{matrix}
		\delta_1 & 0 & \frac{\Omega_1}{2}\\
		0 & \delta_2 & \frac{\Omega_2}{2}\\
		\frac{\Omega_1}{2} & \frac{\Omega_2}{2} & 0
	\end{matrix}
	\right).
	\label{eq:Hamiltonian}
\end{eqnarray}
Considering the relaxation and decoherence phenomenon, the matrix $\mathcal{R}\rho_{ij}$ can be written as,
\begin{eqnarray}
	\mathcal{R}\rho=
	\left(
	\begin{matrix}
		\Gamma_{31}\rho_{33} & 0 & 0\\
		0 & \Gamma_{32}\rho_{33} & 0\\
		0 & 0 & \Gamma\rho_{33}
	\end{matrix}
	\right).
	\label{eq:decoherence}
\end{eqnarray}
Here, the total spontaneous emissivity $\Gamma$ is composed of the rate $\Gamma_{31}$ and $\Gamma_{32}$.
The optical Bloch equation shown in the following formula describes the time evolution of the density matrix element under the RWA,
\begin{eqnarray}
	\frac{d\rho_{11}}{dt} =&& -\Omega_1\rm{Im}(\rho_{13})+\Gamma_{31}\rho_{33},\nonumber\\
	\frac{d\rho_{22}}{dt} =&& -\Omega_2\rm{Im}(\rho_{23})+\Gamma_{32}\rho_{33},\nonumber\\
	\frac{d\rho_{33}}{dt} =&& \Omega_1\rm{Im}(\rho_{13})+\Omega_2\rm{Im}(\rho_{23})-\Gamma\rho_{33},\nonumber\\
	\frac{d\rho_{13}}{dt} =&& -(\frac{\Gamma}{2}+i\delta_1)\rho_{13}+i\frac{\Omega_2}{2}\rho_{12}-i\frac{\Omega_1}{2}(\rho_{33}-\rho_{11}),\nonumber\\
	\frac{d\rho_{23}}{dt} =&& -(\frac{\Gamma}{2}+i\delta_2)\rho_{23}+i\frac{\Omega_2}{2}\rho_{21}-i\frac{\Omega_2}{2}(\rho_{33}-\rho_{22}),\nonumber\\
	\frac{d\rho_{12}}{dt} =&& -i(\delta_1-\delta_2)\rho_{12}+i\frac{\Omega_2}{2}\rho_{13}-i\frac{\Omega_1}{2}\rho_{32}.
	\label{eq:master_equation_Multiple}
\end{eqnarray}

In our experiment, the intensities of the two laser fields are equal.
The corresponding Rabi frequencies are $\Omega_1=\Omega_2=0.106$ MHz and the average Rabi frequency is $\Omega=\sqrt{(\vert\Omega_1\vert^2+\vert\Omega_2\vert^2)/2}$.
To study the pulse CPT-Ramsey scheme, an interrogation time $ T_R$ is inserted into the middle of two CPT pulses, which are called preparation and detection respectively.
Different from the conventional Ramsey interferometry, the CPT pulse is not the typical $\frac{\pi}{2}$-pulse.
The preparation pulse should be long enough to prepare the atoms into the dark state, while the detection pulse should be short enough to avoid repumping the atoms into the dark state again.

The CPT-Ramsey fringe can be detected via the fluorescence signal or the transmission signal of the detection pulse.
In particular, when the decay rate of the excited state is far greater than all other rates in the system, the adiabatic approximation can be applied to the excited state $\vert 3\rangle$.
Therefore an analytical expression of the excited state population at the end of the interaction is given as \cite{Hemmer:89},
\begin{eqnarray}\label{eq:formular_for_rho33}
	&&\rho_{33}(\tau_p+T+\tau_d)= \nonumber \\
 &&\alpha e^{-\alpha\Gamma\tau_d}[1-(1-e^{\alpha\Gamma\tau_p})\vert \rm{sec}(\phi)\vert \rm{cos}(\Delta T_R-\phi_{ls})],
\end{eqnarray}
where $\alpha=\Omega^2/(\Gamma^2+3\Omega^2+4\delta^2)$ represents the interaction with CPT pulses.
The transmission signals $S_{TS}$ , which corresponds to the clock states' coherence, collected during the detection pulse is proportional to $(1-\rho_{33})$,
\begin{eqnarray}
	 S_{TS}\propto\int_{\tau_p+\rm{T_R}}^{\tau_p+\rm{T_R}+\tau_d}[1-\rho_{33}(\tau_d^{\prime})]d\tau_d^{\prime}.
	\label{eq:formular_signals}
\end{eqnarray}

In our experiment, the CPT-Ramsey spectra are constructed from the average PD signals in the detection pulse.
As the detection pulse $\tau_d$ is fixed in the experiment, the transmission signals is  in form of a cosine function $S_{TS}\propto \cos(\Delta T_R-\phi_{ls})=\cos{2\pi(f-f_c + f_s)T_R}$.
Here, the two-photon detuning $\Delta$ is expressed by the LO frequency $f$ and clock frequency $f_c$.
For the Ramsey interrogation, there are residual light frequency shifts induced by the probe field during the interrogation pulses.
These interrogation-related shifts originate from a phase shift $\phi_{ls}$ accumulated during the interrogation pulses and inversely proportional to the interrogation time as $f_s =-\phi_{ls} / (2\pi T_R)$~\cite{Limaojie_2024}.
These $T_R$-dependent frequency shifts is compensated into the LO frequency in our experiment, see Appendix D.

\section{Bayesian quantum parameter estimation for clock frequency}\label{Sec3}

Unlike the frequentist estimation, the Bayesian estimation relies on updating the current knowledge of parameters after each experiment by means of Bayes' law~\cite{Gebhart2023}.
It is particularly suitable for adaptive experiments in which measurement can be optimized based on the current knowledge of parameters.
The adaptivity can improve sensitivity and save time compared with the frequentist estimation.
It has been demonstrated that the sensitivity of Ramsey interferometry via single spin systems can surpass the SQL~\cite{Cappellaro2012, PhysRevX.9.021019,Nusran2012, Bonato2016,Degen2017}.
However, single spin systems can only provide a binary data in each measurement and the efficiency is easily affected by quantum shot noise and decoherence.
Quantum sensors with atomic ensembles have the advantages of larger particle number $N$ and higher signal-to-noise ratio (SNR).
In this section, we present how the Bayesian quantum parameter estimation can be used to enhance the clock frequency estimation.

\begin{figure*}
	\includegraphics[width=1\linewidth]{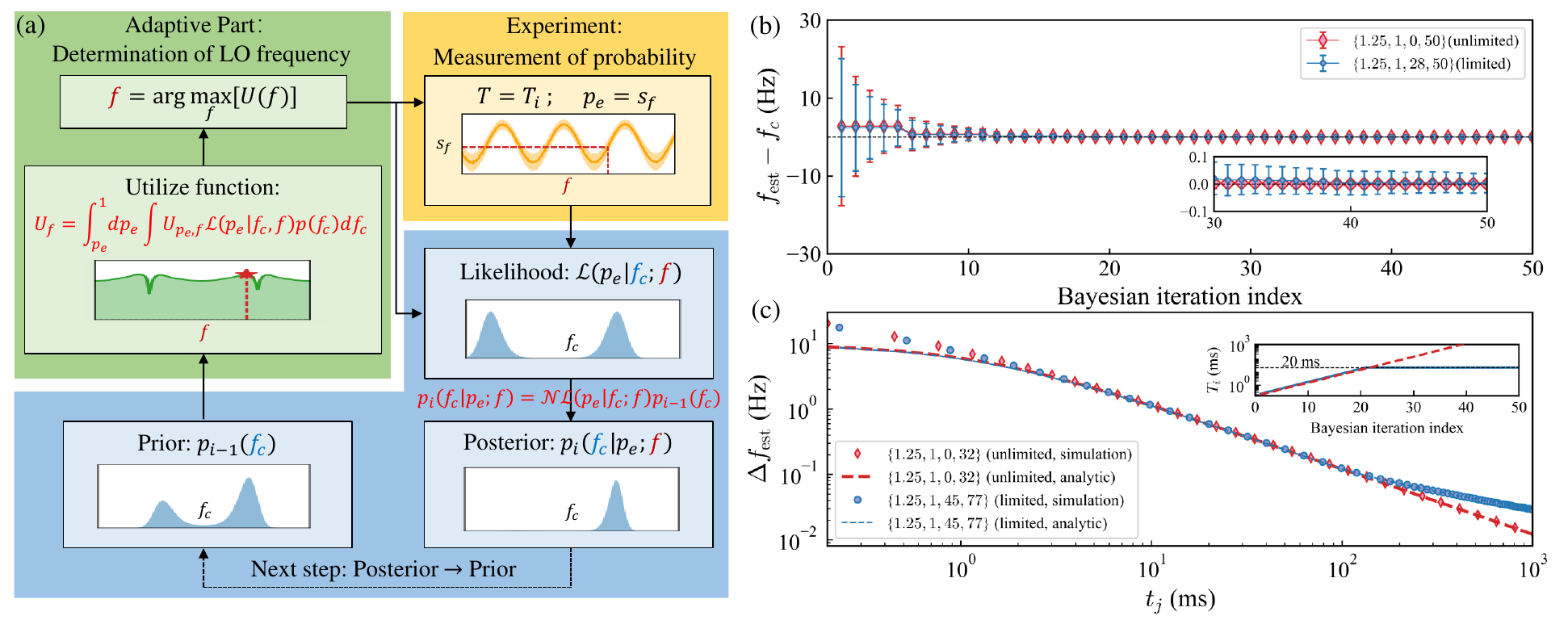}
	\caption{\label{fig_simu}
(a) Schematic of our BQFE protocol. Given a prior distribution $p_{i-1}$ by the previous update, the posterior distribution is updated through the Bayes' formula Eq.~\eqref{Bayes_update}. The likelihood function is obtained with the LO frequency $f$ optimized by the Utilize function Eq.~\eqref{utilize}.
(b) The difference between the estimated and exact frequencies rapidly converges to zero. Red line: the case of increasing $T_i$ without limitation \{$a=1.25, g=1, \tilde M=0, M_b=50$\}. Blue line: the case of increasing $T_i$ under limitation $T_i\le T_{max}$ \{$a=1.25, g=1, \tilde M=28, M_b=50$\}. The errorbars denote the standard deviations.
(c) The standard deviation versus the interrogation time $t_j=\sum_{i=1}^{j} T_i$ for \{$a=1.25, g=1, \tilde M=0, M_b=32$\} and \{$a=1.25, g=1, \tilde M=45, M_b=77$\}, where $\tilde M = M_b - j$. 
Inset: the individual interrogation time $T_i$ versus the number of Bayesian iteration index.
}
\end{figure*}

Below we present the general procedure of Bayesian quantum frequency estimation (BQFE) for determining the clock frequency. 
In a cold-atom CPT clock, the BQFE operates based upon the CPT-Ramsey interferometry. 
The normalized CPT-Ramsey signal $S_{TS}$ can be expressed as
\begin{equation}\label{signal}
  s_f = \frac{1}{2}[1+\cos{2\pi(f-f_c+f_s) T_R}].
\end{equation}
The normalization is feasible as long as more than a period of oscillation can be achieved.
In a single-particle Ramsey interferometry, the likelihood function reads 
\begin{equation}
    \mathcal{L}_u(u\vert f_c, f)= \frac{1}{2}[1+(-1)^{u}\cos{2\pi(f-f_c+ f_s) T_R}], 
\end{equation}
where $u=0$ or $1$ stand for the particle occupying the clock state $\vert F=1,m_F=0\rangle$ or $\vert F=2,m_F=0\rangle$, respectively.
In the CPT-Ramsey interferometry, the signal of each measurement is provided by an ensemble of atoms rather than a single atom.
This implies that the probability ($p_e$) of the atoms occupying the clock state follows a binomial distribution, which can be approximated by a Gaussian distribution when the total particle number in the experiment is sufficiently large.
Thus we use a Gaussian distribution function as our likelihood function~\cite{Dinani2019,Craigie2021},
\begin{equation}\label{likelihood_ensemble}
    \mathcal{L}(p_e\vert f_c, f)= \frac{1}{\sqrt{2\pi}\sigma}\exp{\left[-\frac{(p_e-\mathcal{L}_u(1\vert f_c,f))^2}{2\sigma^2}\right]},
\end{equation}
where $\sigma^2  \approx p_e(1-p_e)/R = s_f(1-s_f)/R $ with $s_f$ the normalized Ramsey signal and $R$ determined by the signal's SNR. Here, $R=1540$ in our experiment, see Appendix B.
This likelihood $\mathcal{L}(p_e\vert f_c, f)$ is a function of $p_e$ with respect to LO frequency $f$. 

In the Bayesian update, the probability distribution is updated according to the Bayes' formula,   \begin{equation}\label{Bayes_update}
    p_i(f_c\vert p_e;f) = \mathcal{N} \mathcal{L}(p_e\vert f_c, f) p_{i-1}(f_c),
\end{equation}
where $\mathcal{N}$ is a normalization factor and the initial prior function $p_1(f_c)$ can be chosen as a uniform distribution function over the measurement range of $f_c$, see Algorithm~\ref{BPE_algorithm} in Appendix A.
In the $i$-th update, we obtain a posterior function $p_i(f_c\vert p_e;f)$,
and the estimator of $f_c$ can be given as $f_{\rm{est}}^{(i)} = \int{f_c p_i\left(f_c\vert p_e;f\right) {\rm d} f_c}$  with a standard deviation $\Delta f_{\rm{est}}^{(i)} = \sqrt{ \int{f_c^2 p_i(f_c\vert p_e;f)  {\rm d} f_c} - \left(f_{\rm{est}}^{(i)}\right)^2}$.
The next update is implemented by inheriting the posterior function as the next prior function $p_{i+1}(f_c)=p_i(f_c\vert p_e;f)$.

For the periodic probability distribution and the Gaussian likelihood, %~\eqref{likelihood_ensemble}
one may use the adaptive strategy with exponentially sparse interrogation times~\cite{Said2011,Cappellaro2012,Nusran2012,Waldherr2012,Bonato2016,articleFerrie}.
To estimate the clock frequency, we adaptively adjust the LO frequency  according to the posterior distribution during Bayesian updates.
The procedure of BQFE protocol is sketched in Fig.~\ref{fig_simu}~(a).
%
%At begin, the interrogation time $T_1$ and the LO frequency $f$ should be preset.
Here, we denote the interrogation time in the $i$-th Bayesian update as $T_R^{(i)} \equiv T_i$.
In practice, $T_i$ is limited by the coherence time $T_{max}$. 

To design an optimal exponential sequence of $T_i$, we set the last interrogation time $T_{M_b}=T_{max}$ and derive the previous ones as 
\begin{equation}\label{exp_seq}
    T_i = T_{max}/a^{M_b-i},
\end{equation}
with $a>1$.
The number of iterations $M_b=\log_a(T_{max}/T_{min})+1$ can be determined by the ratio between maximum and minimum interrogation time $T_{max}$ and $T_{min}$.
In our experiment, $T_i$ is limited up to $T_{max}=20 \rm $ ms due to the freely falling atoms will be out of the CPT beam, and the available minimum interrogation time $T_{min}\ge0.2$ ms. Thus, $T_i$ should satisfy $ T_{min} \le T_i \le T_{max}$. 
Since the first interrogation time is determined by $T_{max}$ and $a$, the initial interrogation time is often given as $T_1=\textrm{max}\{T_{min}, T_{max}/a^{M_b-1}\}$.

Therefore, we have the total interrogation time 
\begin{equation}
    T=\sum_{i=1}^{M_b}T_i\approx T_{max} \frac{a}{a-1}.
\end{equation}
The initial probability distribution of $f_c$ is set in the interval $[f_l, f_r]$ of width $f_{lr}\equiv f_r-f_l=1/T_1$.
After each update via Eq.~\eqref{Bayes_update}, the corresponding frequency range interval is turned into $[f_{\rm{est}}^{(i)}-1/2T_i, f_{\rm{est}}^{(i)}+1/2T_i]$ as $T_i$ increases.

In the adaptive procedure, the LO frequency $f_i$ for the $i$-th update is determined with the previous posterior distribution $p_{i-1}(f_c|p_e; f)$.
The likelihood function Eq.~\eqref{likelihood_ensemble} is updated adaptively so that the spurious peaks in the posterior distributions can be eliminated~\cite{Lumino2018}.
To give $f_i$, we use the expected gain in Shannon information of the posterior distribution~\cite{Ruster2017}.
Different from the summation over single-shot measurements~\cite{Ruster2017}, we measure an ensemble of particles and the Utilize function becomes
\begin{equation}\label{utilize}
    U_f = \int_{0}^1{\rm d}p_e {\int{U_{p_e,f}\mathcal{L}(p_e\vert f_c, f)p(f_c){\rm d}f_c}},
\end{equation}
where 
\begin{eqnarray}
    U_{p_e,f} &=& \int{p(f_c\vert p_e;f)\ln{[p(f_c\vert p_e;f)]}{\rm d}f_c} \nonumber \\
    &-&\int{p(f_c)\ln{[p(f_c)]}{\rm d}f_c},    
\end{eqnarray}
is the expected gain in Shannon information of the posterior distribution with respect to the prior distribution.
Thus $f_i$ is chosen as the one that maximizing $U_{f_i}$.
For efficiency, we calculate the Utilize function by discretizing the integral over $p_e$, see Appendix B.

\begin{figure}[htp]
    \includegraphics[width = 1\columnwidth]{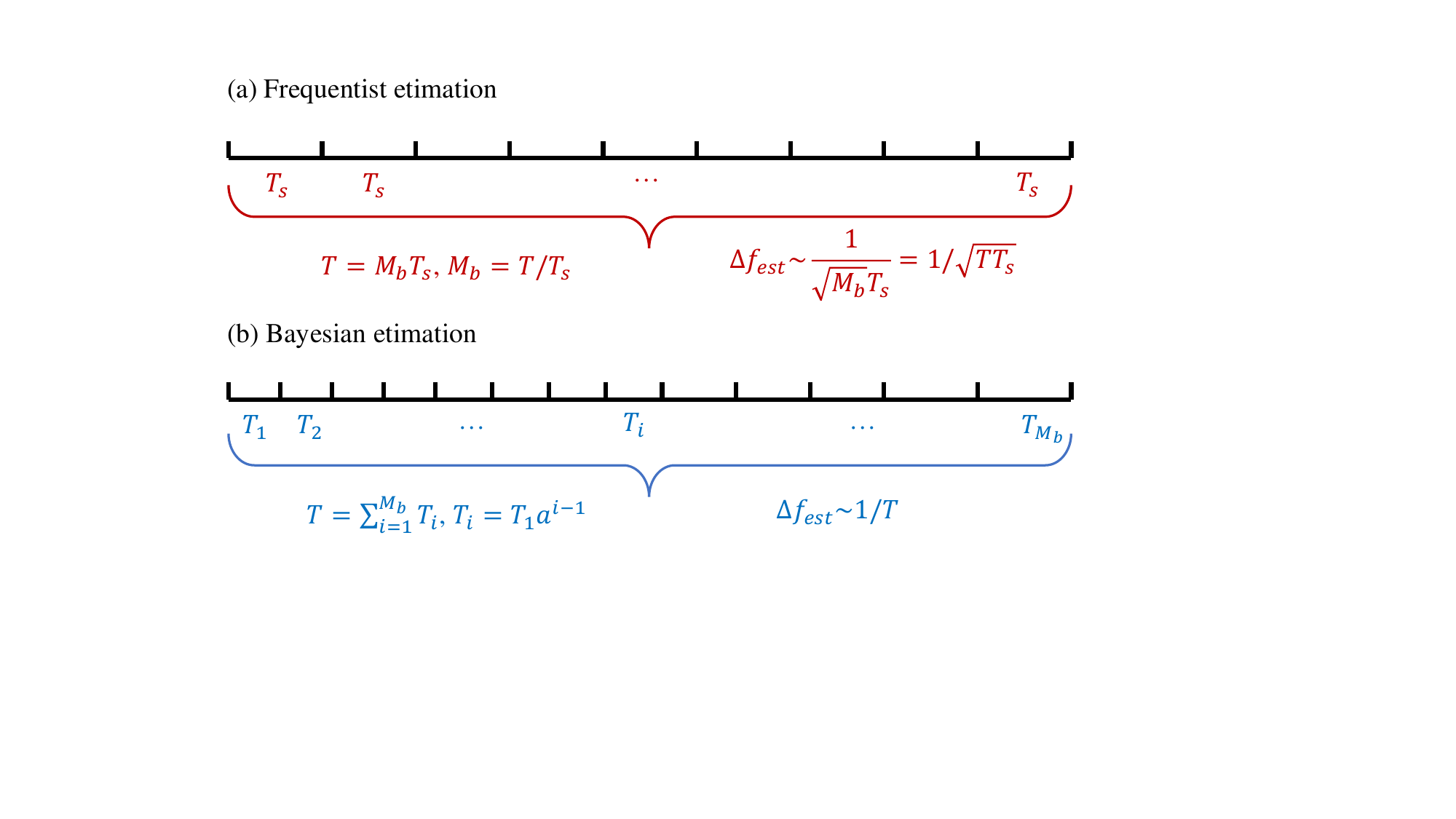}
    \caption{\label{frequency_scaling}
    Schematics of interrogation time sequences for (a) conventional frequentist estimation and (b) Bayesian quantum estimation.
    }
\end{figure}

Once $f_i$ and $T_i$ are given, a measurement can be made to obtain a normalized signal $s_{f_i}$.
Due to the $T_i$-dependent frequency shift $f_s$, a measurement of $f_s$ prior to the Bayesian update should be made for a compensation, see Appendix D.
In Fig.~\ref{fig_simu}, we show the numerical performance of our BQFE protocol.
By setting $a=1.25$, the estimated value $f_{\rm est}$ is gradually converged to the clock frequency $f_c$.
The scaling of the standard deviation 
\begin{equation}
    \Delta f_{\rm est} \approx \sqrt{1+\frac{2}{a-1}}\frac{1}{2\pi\sqrt{R} T}
\end{equation}
versus the total interrogation time $T=a T_{max}/{(a-1)}$ obeys the Heisenberg scaling $ \Delta f_{\rm est} \propto T^{-1}$.
%
%Since the scaling is inversely proportional to the total interrogation time $T$ rather than a single Ramsey interrogation time $T_{s}$, it holds for $T \gg T_{s}$, which gives a $\sqrt{T}$-fold enhancement of the SQL scaling achieved by conventional frequentist estimation.

\subsection{Heisenberg scaling with respect to the total interrogation time}

%In a conventional frequentist estimation of $T/T_{s}$ times of independent measurements over a fixed interrogation time $T_{s}$, which always satisfies $T_{s}\ll T$, the standard deviation obeys the SQL scaling $\propto 1/\sqrt{T T_{s}} \propto 1/\sqrt{T}$.  
%
%It should emphasize that for the special case where only one measurement is performed with $T=T_{s}$, the standard deviation would be $\propto 1/\sqrt{T T_{s}} \propto 1/T$. 
%
%However, this scaling will not hold when $T>T_s$. 
%
%In contrast, in our Bayesian estimation procedure, we perform a sequence of correlated measurements with increasing interrogation times ${T_i}$ over the whole duration $T= \sum_{i=1}^{M_b} {T_i}$, in which the parameters for the next measurement are determined by its previous measurements.
%Therefore the standard deviation of our Bayesian estimation procedure attains the Heisenberg scaling $\propto 1/T$ during $T$~\cite{SM}.
%

The SQL is originated from the multiple individual measurements made over a short interrogation time. 
Take the conventional frequentist scheme for an example. If one repeats the Ramsey interferometry with the interrogation time $T_R$ for $M_b$ times.
Without loss of generality, we assume the Ramsey interrogation time equals $T_R=T_s$, which is not greater than the coherence time $T_{max}$.
For total interrogation time $T$ with $T \gg T_s$, we can perform $M_b=T/T_s$ times of Ramsey interferometry, as shown in Fig.~\ref{frequency_scaling}~(a).
Thus according to the central limit theorem, we have $\Delta f_{est} \propto \frac{1}{T_s\sqrt{M_b}} = 1/\sqrt{T T_s}\propto 1/\sqrt{T}$, which is the SQL.
Only when $T=T_{s}$, the standard deviation would be $\propto 1/\sqrt{T T_{s}} \propto 1/T$. 
This is a special case and it will break down when $T>T_s$. 
Thus we have the fact that the scaling versus $T$ is just the SQL for a total interrogation time $T \gg T_s$.  

While in our protocol, we use BQFE algorithm with a sequence of interrogation times $T_i$ for Ramsey interferometry, as shown in Fig.~\ref{frequency_scaling}~(b).
Updating the probability distribution via Bayesian approach, we demonstrate that the sensitivity obeys the Heisenberg scaling $\Delta f_{est} \propto 1/T$ versus the total interrogation time $T= \sum_{i=1}^{M_b} {T_i}$.
Since the scaling is inversely proportional to the total interrogation time $T$ rather than a single Ramsey interrogation time $T_{s}$, it holds for $T \gg T_{s}$, which gives a $\sqrt{T}$-fold enhancement of the SQL scaling achieved by conventional frequentist estimation.

This Heisenberg scaling can hold when $T>T_{max}$, and it can be analytically explained by using Bayes' rule with multiple Gaussian probability distributions iteratively~\cite{Craigie2021}. 
Since the measured $p_e$ in the experiment is related to the probability $\mathcal{L}_u(1\vert f_c,f)$, we have $p_e\approx\mathcal{L}_u(1\vert f_c,f)=\frac{1}{2}[1-\cos{2\pi(f-f_c+ f_s) T_R}]$.
Then, we get $2\pi(f-f_c+ f_s) T_R \approx \Phi$ with $\Phi=\pm \arccos(1-2p_e) + k\pi$ ($k$ is an integer).
In this case, we can perform the Taylor expansion around $\Phi$, 
\begin{widetext}
\begin{eqnarray}\label{Taylor}
\frac{1}{2}[1-\cos{2\pi(f-f_c+ f_s) T_R}]  \approx 
\frac{1}{2}\left\{1-\left[\cos \Phi -\sin \Phi (2\pi(f-f_c+ f_s) T_R-\Phi)\right]\right\}.
\end{eqnarray}

Substituting Eq.~\eqref{Taylor} into Eq.~\eqref{likelihood_ensemble}, the likelihood function can be calculated as
\begin{equation}\label{likelihood_f}
    \mathcal{L}(p_e\vert f_c, f)= \frac{1}{\sqrt{2\pi}\sigma} \exp{\left[-\frac{(p_e-\frac{1}{2}\left\{1-\left[\cos \Phi -\sin \Phi (2\pi(f-f_c+ f_s) T_R-\Phi)\right]\right\})^2}{2\sigma^2}\right]}.
\end{equation}
\end{widetext}
After some algebra, the normalized likelihood function versus $f$ can be simplified as 
\begin{equation}\label{likelihood_f_Gaussian}
    \mathcal{L}(f)= \frac{1}{\sqrt{2\pi}\sigma_f} \exp{\left[-\frac{(f-f^{\prime})^2}{2\sigma_f^2}\right]},
\end{equation}
which is also in the form of Gaussian function in the range of $f\in[f^{\prime}-3\sigma_f,f^{\prime}+3\sigma_f]$ with $f^{\prime}=f_c-f_s+(p_e-\frac{1}{2}+\frac{1}{2}\cos\Phi+\frac{1}{2}\Phi \sin\Phi)/(\pi T_R \sin \Phi)$ and $\sigma_f=\sigma/(\pi T_R \sin \Phi)$.
Generally, for our BQFE, the probability $p_e$ will gradually converge to $p_e\approx \frac{1}{2}$ after several iterations. 
In this case for simplicity, we first assume $p_e=\frac{1}{2
}$ with $\Phi=\frac{\pi}{2}$ for all iterations. 

Thus, the likelihood is always in the form of Gaussian function as Eq.~\eqref{likelihood_ensemble},
\begin{equation}\label{Gaussian}
    \mathcal{L}_i=\frac{1}{\sqrt{2\pi} \sigma_i}\exp{\left[-\frac{(f-\mu_i)^2}{2\sigma_i^2}\right]}.
\end{equation}
Here, the uncertainty of the Gaussian function $\sigma_i = C/T_i  \propto  1/T_i$ with a constant $C=1/(2\pi\sqrt{R})$.
Then, the final posterior probability after $M_b$ times of Bayesian updates is equivalent to the product of $M_b$ multiple Gaussian functions, which can be written as
\begin{equation}\label{Gaussian2}
    \mathcal{N} \prod_{i=1}^{M_b} \mathcal{L}_i=\mathcal{N} \prod_{i=1}^{M_b} \left[\frac{1}{\sqrt{2\pi} \sigma_i} \exp{\left(-\frac{(f-\mu_i)^2}{2\sigma_i^2}\right)}\right],
\end{equation}
where $\mathcal{N}$ is the normalized factor.

The product of the first two Gaussian functions can be analytically calculated as~\cite{PhysRevA.76.033613}
\begin{widetext}
\begin{equation}\label{Gaussian3}
    \left[\frac{1}{\sqrt{2\pi} \sigma_1}\exp{\left(-\frac{(f-\mu_1)^2}{2\sigma_1^2}\right)}\right] \left[\frac{1}{\sqrt{2\pi} \sigma_2}\exp{\left(-\frac{(f-\mu_2)^2}{2\sigma_2^2}\right)}\right] = \frac{A}{\sqrt{2\pi} \sigma^{\prime}}\exp{\left(-\frac{(f-\mu^{\prime})^2}{2{\sigma^{\prime}}^2}\right)},
\end{equation}
\end{widetext}
where $\mu^{\prime}=\frac{\mu_2 \sigma_1^2 + \mu_1 \sigma_2^2}{{\sigma_1^2+\sigma_2^2}}$, $\sigma^{\prime}=\frac{\sigma_1 \sigma_2}{\sqrt{\sigma_1^2+\sigma_2^2}}$ and $A$ is a scale factor.
Then the normalized posterior for the second step can be approximated as
\begin{equation}\label{Gaussian4}
    p_{2}=\frac{1}{\sqrt{2\pi} \sigma^{\prime}}\exp{\left(-\frac{(f-\mu^{\prime})^2}{2{\sigma^{\prime}}^2}\right)}.
\end{equation}
The location of the center frequency $\mu^{\prime}=\frac{\mu_2 \sigma_1^2 + \mu_1 \sigma_1^2/a^2}{{\sigma_1^2+\sigma_1^2/a^2}}=\frac{\mu_1+\mu_2 a^2}{1+a^2}$.
Obviously,  if $a$ near 1, $\mu'$ is closer to $\mu_1$, the frequency can converge more smoothly and the corresponding dynamic range is high. 
The standard deviation of the frequency $\sigma^{\prime}=\frac{\sigma_1 \sigma_2}{\sqrt{\sigma_1^2+\sigma_2^2}}=\frac{(C/{T_1})(C/{T_2})}{\sqrt{(C/{T_1})^2+(C/{T_2})^2}}=\frac{C}{\sqrt{T_1^2+T_2^2}}$.

\begin{figure*}[ht]
\includegraphics[width=0.85\linewidth]{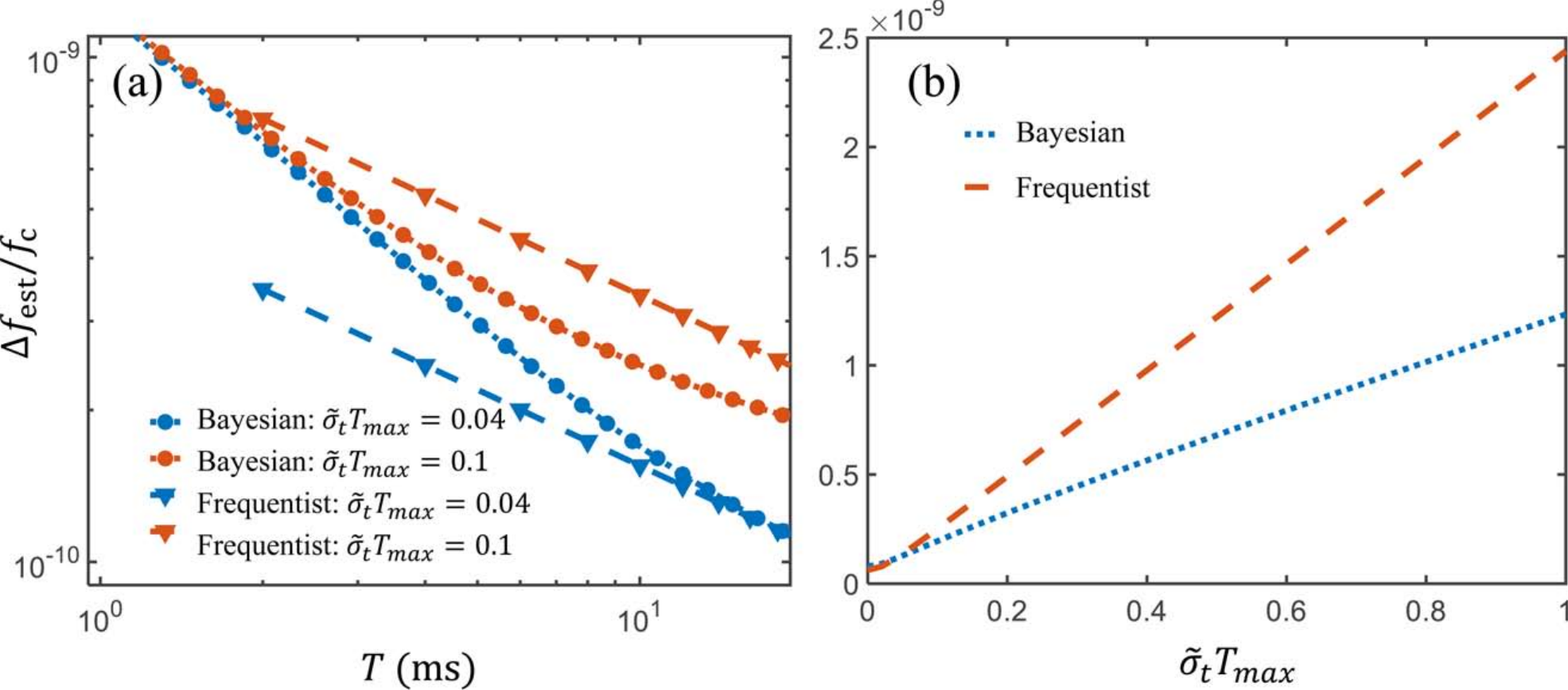}
\caption{\label{fig:robustness}
Robustness against technical noises.
  (a) The blue and orange circles are the simulation results using BQFE under $\tilde \sigma_t T_{max}=0.04$ and $\tilde \sigma_t T_{max}=0.1$, respectively. We choose the exponential interrogation time sequence as Eq.~\eqref{exp_seq} with $a=10/9$ from $T_{min}=0.2$ ms to $T_{max}=20$ ms.  While the blue and orange triangles are the simulation results using frequentist measurement under $\tilde \sigma_t T_{max}=0.04$ and $\tilde \sigma_t T_{max}=0.1$, respectively. We choose $T_s=T_{max}$. (b) The final $\Delta f_{est}/f_c$ with total interrogation time $T\approx 200$ ms versus the technical noise strength using BQFE and frequentist measurement. Here, $R=1540$.}
\end{figure*}

Analytically, one can easily find that the uncertainty of the posterior probability is only dependent on the uncertainties of the prior and likelihood functions.
Since we only focus on the uncertainty in the following, we can perform the product of two Gaussian functions step by step and we eventually get the final posterior probability 
\begin{equation}\label{Gaussian5}
    p_{M_b}=\frac{1}{\sqrt{2\pi} \sigma}\exp{\left(-\frac{(f-\mu)^2}{2\sigma^2}\right)},
\end{equation}
where $\mu \approx f_c$, $\sigma=\frac{C}{\sqrt{\sum_{i=1}^{M_b} T_i^2}}$.

In our scheme, $T_i=T_1 a^{i-1}$ increases exponentially from $T_1$ to $T_{max}$, so we have 
\begin{equation}
    \sigma=\frac{C}{T_{max}} \sqrt{\frac{1-a^{-2}}{1-a^{-2M_b}}}. 
\end{equation}
If $a^{-M_b} \ll 1$, $\sigma \approx \frac{C}{T_{max}} \sqrt{{1-a^{-2}}}$.
Since $T=\sum_{i=1}^{M_b} T_i=T_{max} (1-a^{-M_b})/{(1-a^{-1})}\approx T_{max}\frac{a}{a-1}$, and we can get the Heisenberg scaling 
\begin{equation}\label{HL}
    \sigma \approx \frac{C}{T} \sqrt{1+2/(a-1)} \propto 1/T.
\end{equation}
The total interrogation time $T$ is $\frac{a}{a-1}$ times larger than $T_{max}$.
Thus using the BQFE protocol, the Heisenberg-limited sensitivity can be achieved for a time $T$ much longer than the coherence time $T_{max}$, see Fig.~\ref{fig_simu}~(b) and (c).

%Moreover, our BQFE has much better robustness against experimental noises~\cite{SM}. 

\subsection{Robustness against technical noises}

Our BQFE protocol shows high robustness against technical noises. 
Assuming the other technical noises can be characterized by a stochastic Gaussian function 
\begin{equation}
    \tilde{\mathcal{N}}_t = \frac{1}{\sqrt{2\pi} \tilde{\sigma}_t}\exp{\left[-\frac{(f-\mu_i)^2}{2\tilde{\sigma}_t^2}\right]},
\end{equation}
where $\tilde \sigma_t$ is the strength of the technical noise. 
Since the quantum projection noise and the technical noise are independent, the variance of the likelihood function Eq.~\eqref{Gaussian} becomes 
\begin{equation}\label{Gaussian-noise}
    \mathcal{L}_i=\frac{1}{\sqrt{2\pi} \tilde{\sigma}_i}\exp{\left[-\frac{(f-\mu_i)^2}{2\tilde{\sigma}_i^2}\right]},
\end{equation}
with $\tilde \sigma_i = \sqrt{\sigma_i^2 +\tilde{\sigma}_t^2}$.
Still by multiplying the Gaussian functions step by step one can get the final posterior probability.
As expected, the final standard deviation would become large as $\tilde \sigma_t$. 

Interestingly, when $T_i$ is short, the quantum projection noise $\sigma_i=C/{T_i}$ is large. 
If $\sigma_i$ can submerge $\tilde \sigma_t$, i.e.,
$\sigma_i \gg \tilde \sigma_t$, 
\begin{equation}
    \tilde \sigma_i = \sqrt{\sigma_i^2 +\tilde{\sigma}_t^2} \approx \sigma_i=\frac{C}{T_i},
\end{equation}
the technical noise will hardly influence the likelihood function. 
Thus, at the beginning with short interrogation times, the quantum projection noise may be larger than the technical noise and the standard deviation of the frequency will be close to the ideal one. 
Thus, our BQFE protocol can be robust against other experimental noises, see Fig.~\ref{fig:robustness}~(a).
However, the effect of $\tilde \sigma_t$ will accumulate iteratively. 
When $T_i$ gets large enough that $\sigma_i \sim \tilde \sigma_t$, the technical noise begins to decrease the measurement precision significantly. 
In general, in the presence of noises, one can choose a smaller $a$ so that $T_i$ increases slower, in order to enhance the robustness of the BFQE. 
In addition, the ways of increasing $T_i$ are diverse and we will show various routines for achieving the robust high-precision estimation in Sec.~\ref{Schemes}.

In comparison, the standard deviation using frequentist measurement under technical noise is $\Delta f_{est}=\frac{\sqrt{C^2+\tilde \sigma_t^2 T_s^2}}{\sqrt{T T_s}}$.
When the quantum projection noise is small with long $T_s$, the influences of technical noise will become significant and decrease the measurement precision dramatically. 
The final measurement precision versus the technical noise strength for the two methods are shown in Fig.~\ref{fig:robustness}~(b).
%
%Another feature of robustness of our BQFE is that it can be immune from the sudden change of the atomic resonance frequency induced by some unknown external field, which will be illustrated in Sec.~\ref{robust_BQFE}.

\begin{figure*}[ht]
\includegraphics[width=1\linewidth]{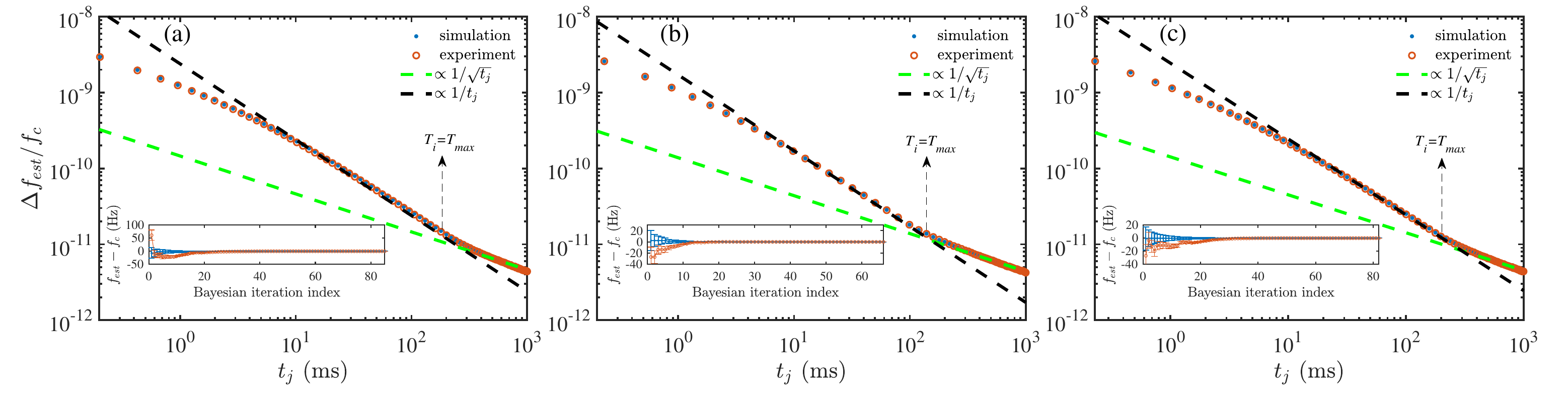}
\caption{\label{fig_exp_resluts}
  Experimental demonstration of Bayesian quantum frequency estimation.
  The standard deviation versus the interrogation time $t_j=\sum_{i=1}^{j} T_i$: (a) \{$a=10/9, g=1,\tilde M=40, M_b=85$\}, (b) \{$a=1.25, g=1, \tilde M=44, M_b=66$\}, and (c) \{$a=1.25, g=2, \tilde M=37, M_b=81$\}.
  Insets: the estimated frequency bias versus the Bayesian iteration index with the corresponding standard deviations denoted by errorbars.
  The numerical results are also provided for comparison.
  }
\end{figure*} 

\subsection{Schemes for achieving robust and high-precision estimation}\label{Schemes}

Given a total interrogation time $T$, there are various routines to improve measurement precision by employing Bayesian quantum parameter estimation.

First, choose the last $T_i$ as $T_{max}$ and change $T_i$ every $g$ steps, that is, $T_{M_b}=T_{max}$ and $T_i = T_{max}/ a^{\lceil (M_{b}-i)/g \rceil}$ with $a=T/(T-gT_{max})$ and $g$ being a positive integer. 
Thus we have the total iteration number $M_b=g \left[\log_{a}(T_{max}/T_{min})+1\right]$ and the final measurement precision is given as 
\begin{equation}
    \sigma\approx\frac{C}{\sqrt{\sum_{i=1}^{M} T_i^2}}=\frac{C}{\sqrt{g \sum_{i=1}^{M_{\tilde b}} T_i^2}}=\frac{C}{T_{max}} \sqrt{\frac{1-a^{-2}}{g}}.
\end{equation}

Second, when the interrogation time $T_{i}$ attains the coherence time $T_{max}$, $T_i$ ($i\ge j$) cannot increase any more in practice.
Thus one can perform the BQFE continuously with $T_{i} = T_{max}$ for $j \le i \le M_{b}$.
In this scenario, $T_i$ increases exponentially from $T_1$ to $T_{max}$ within the first $j$ steps, and subsequently $T_i$ stays constant at $T_{max}$ for the remaining $\tilde M = (M_b-j)$ steps.
Then the total interrogation time $T=T_{max}(\frac{a}{a-1}+\tilde M)$, where $\tilde M = M_b - j$.
The measurement precision becomes
\begin{equation}
    \sigma=\frac{C}{T_{max}}\sqrt{\frac{1}{\tilde M+\frac{a^2}{a^2-1}}}.
\end{equation} 
In particular when $\tilde M \gg \frac{a}{a-1}$, $T \approx \tilde M T_{max}$, it reduces  to $\sigma \approx \frac{C}{\sqrt{T T_{max}}}\propto 1/\sqrt{T}$, which is the SQL scaling when $T$ is large.

Third, combing the previous two schemes, in which the $i$-th interrogation time can be expressed as
\begin{equation}\label{T_increase}
	T_i=\begin{cases}
	T_{\rm max}/ a^{\beta_i}, &  
 \beta_i \in \mathbb{N^+} \textrm{~and~} i < M_b-\tilde M,
 \\
	T_{i-1}, &  \beta_i \notin \mathbb{N^+} \textrm{~and~} i < M_b-\tilde M,\\
        T_{\rm max}, &  i \ge M_b-\tilde M,
	\end{cases}
\end{equation}
where the exponent $\beta_i=(M_b-\tilde M-i)/g$.
For convenience, we label all above schemes with $\{a,g,\tilde M, M_b\}$ with more details shown in Appendix C.

We use the BQFE protocol to measure the clock frequency of $^{87}{\rm Rb}$ atom via CPT in a cold atomic ensemble.
The experiment is performed automatically with the help of a computer connecting with digital I/O devices.
The BQFE protocol used in our experiment is executed individually to the numerical simulation, except the parameter $p_e$ that is dependent on the normalized Ramsey signals $s_f$.
By choosing $T_1=0.2$ ms, without loss of generality, we set the frequency range $[f_l=f_c-3~{\rm kHz}, f_r=f_c+2~{\rm kHz}]$ as the initial interval.
As $T_1$ and $a$ are preset, we can compensate the $T_i$-dependent frequency shift $f_s$ via adding it to the LO frequency as $f^{\prime}_i=f_i+ f_s$.
Then two varying parameters $T_i$ and $f^{\prime}_i$ in each update can be obtained via time sequence and Eq.~\eqref{utilize}, respectively.
These two parameters are sent to the digital I/O devices that control a FPGA to generate a CPT-Ramsey sequence and scan the LO frequency to acquire the CPT-Ramsey fringe.
The normalized signal $s_{f_i}$ is extracted by cosine fitting of the CPT-Ramsey fringe.
The LO is referenced to a high-performance rubidium atomic clock with stability of $6\times10^{-13}$ at 1 s.

The clock frequency $f_c$ are measured with three different Bayesian update schemes: \{$a=10/9, g=1,\tilde M=40, M_b=85$\}, \{$a=1.25, g=1, \tilde M=44, M_b=66$\} and \{$a=1.25, g=2, \tilde M=37, M_b=81$\}.
When the iteration number is small, the experimental $f_{est}$ deviate from numerical ones due to the influence of experimental noises (see insets of Fig.~\ref{fig_exp_resluts}).
However, it gradually converges to an acceptable value within $1\sigma$ standard deviation as the Bayesian iteration number increases.
Meanwhile the standard deviations provided by the BQFE agree well with the numerical and analytic ones, see Fig.~\ref{fig_exp_resluts}.
Notably, the fractional measurement precision $\Delta f_{est}/f_c$ follows the Heisenberg scaling before $T_i$ reaches $T_{max}$.

Besides, our BQFE exhibits high dynamic range. 
In conventional frequentist estimation, one may choose $T_{s}=T_{max}$ to achieve the highest precision, but the corresponding dynamic range would become small due to the phase ambiguities~\cite{Said2011,Waldherr2012,Nusran2012}.
However, for our BQFE, its dynamic range is mainly determined by the minimum interrogation time $T_1$ in the measurement sequence and its measurement precision can be improved without sacrificing the dynamic range. 
More details of dynamic range can be found in Appendix C.

\section{Closed-loop locking of cold-atom CPT clock}\label{Sec4}

In this section, we show how to use the Bayesian quantum frequency estimation (BQFE) to enhance the closed-loop locking of a cold-atom CPT clock.
To give a benchmark, we implement the conventional proportional-integral-differential (PID) locking at first.
Then, under the same conditions, we employ the developed BQFE to lock the cold-atom CPT clock. 
Our experimental results demonstrate that, compared to the conventional PID locking, our Bayesian locking not only improves stability by 5.1(4) dB but also exhibits better robustness against technical noises.

\begin{figure*}[ht]
    \includegraphics[width=1\linewidth]{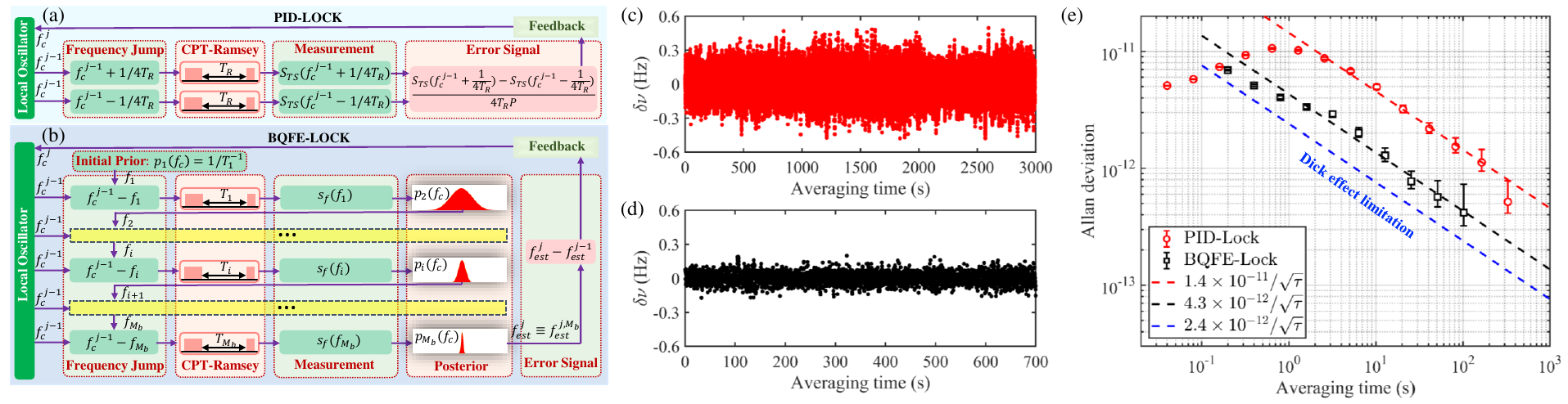}
    \caption{\label{fig_exp_clock} 
	(a) Locking a cold-atom CPT clock via conventional PID controller.
    Two measurements of $S_{TS}$ at bilateral half-maximum points ($f_c^{j-1}+1/4T_R$ and $f_c^{j-1}-1/4T_R$) is used to derived the error signal and fed back to the LO.
	(b) Locking a cold-atom CPT clock via BQFE. 
    $13$ Bayesian iterations are used to estimate the clock frequency $f_c$ and no feedback is provided to the LO until completion of all Bayesian iterations. 
	(c) The frequency fluctuations for PID locking.
	(d) The frequency fluctuations for BQFE locking.
    (e) Allan deviations.
	The red dot and dashed line are the results from PID locking, while the black square and dashed line are the results from BQFE locking.
    The Dick effect limitation (the blue dashed line) is estimated according to the sensitivity function and the LO phase noises.
    }
\end{figure*}

\subsection{Conventional proportional-integral-differential locking}

For pulse-mode atomic clock, the conventional method to estimate the clock frequency is alternate measurements of the transition probability at bilateral half-maximum points of atomic resonance line.
Then the clock frequency is derived from the transition probability difference at these two set points and fed back to stabilize the LO frequency by a PID controller from cycle to cycle.
For Ramsey fringes with a linewidth of $1/2T_R$, these two set points are $f_c+1/4T_R$ and $f_c-1/4T_R$.

In our experiment, a CPT-Ramsey fringe with the Ramsey time $T_R=20$ ms is used for conventional PID locking of the cold-atom CPT clock.
For the $j$-th feedback cycle, the clock frequency $f_c^j =f_c^{j-1}+\kappa \Delta \nu_j$, where $\kappa$ is the servo gain of PID controller.
The error signal $\Delta \nu_j =\frac{S_{TS}(f^{j-1}_c + 1/4 T_R)-S_{TS}(f^{j-1}_c - 1/4 T_R)}{4T_R P}$, where $P$ is the amplitude of CPT-Ramsey central fringe $S_{TS}$ [see Fig.~\ref{fig_exp_clock}~(a)].
By feeding back this error signal to the LO through a PID controller, its frequency is locked to central Ramsey fringe.
As two measurements of $S_{TS}$ in one feedback, the total interrogation time of a feedback $T=40$ ms.
To compare the conventional PID locking with the Bayesian locking on the same scale, the dead time ignored.
The fluctuation $\delta \nu_j =f_c^{j}-f_c$ and the corresponding Allan deviations are shown in Fig.~\ref{fig_exp_clock}~(c) and (e), which yields a stability of $1.4(1)\times10^{-11}/\sqrt{\tau}$.

\subsection{High-precision locking via Bayesian quantum frequency estimation}\label{robust_BQFE}

Using our BQFE protocol, we stabilize the LO frequency to the atomic reference and realize closed-loop locking of a cold-atom CPT clock.
The BQFE procedure is carried out to acquire the clock frequency and directly feed back to the LO frequency via the error signal $\Delta \nu_j =f_{est}^{j} -f_{est}^{j-1}$. 
Here $f_{est}^{j}$ is the $j$-th estimated value of clock frequency obtained by $M_{b}$ Bayesian updates, see Fig.~\ref{fig_exp_clock}~(b).
For the next feedback, the estimated value $f_{est}^{j}$ is inherited as the clock frequency $f_c$, then we reset the BQFE procedure via initializing the frequency interval $[f_l, f_r]$ and prior distribution $p_1(f_c)$ as uniform distribution.
This means that we need $M_{b}$ measurements of $s_{f_i}$ to achieve one feedback rather than two measurements of $S_{TS}$ in conventional locking.
We choose the BQFE of \{$a=1.25$, $g=1$, $\tilde M=6$, $M_{b}=13$\} for performing closed-loop locking. 
By increasing $T_i$ from the initial interrogation time $T_1 = T_{max}/1.25^{M_b-\tilde M-1}$ ms to the maximum interrogation time $T_{max}=20$ ms, we realize a locking of the cold-atom CPT clock.
The total interrogation time $T$ is about 199 ms, which is about $5$ times longer than conventional locking.
Notably, the frequency fluctuation $\delta \nu_j =f_{est}^{j} -f_c$ of BQFE locking is about $3$ times smaller than that of conventional locking with a PID controller, see Fig.~\ref{fig_exp_clock}~(c) and (d).

To assess the clock stability, we analyze the Allan deviations for different locking schemes.
As the transmission signal $S_{TS}$ is proportional to the excited-state population rather than the clock-state population, the normalized signal $s_{f}$ is used for BQFE.
In our experiment, $s_{f}$ is obtained by scanning the CPT-Ramsey fringe, and so that we expend more time to measure $s_{f_i}$ than $S_{TS}$.
However, for ion trap clocks, fountain clocks and optical lattice clocks, the transition probability at LO frequency $f_i$ could be directly probed by the electron shelving technique~\cite{Dehmelt1986,Wynands2005fountain,RevModPhys.87.637} and the corresponding measurement times will decrease greatly.
Anyway, the clock cycle time (a feedback time) $T_c$ contains the total interrogation time $T$ and dead time $T_d$.
Then the averaging time can be expressed as $\tau=M_c T_c=M_c (T+T_d)$, where $M_c$ is the total feedback number.
Here, we use a total interrogation time of a feedback cycle $T=199$ ms for BQFE locking while $T=40$ ms is used for the conventional locking.
Due to the indirect detection of $s_f(f_i)$, the dead time is different for the two locking methods.
In the clock locking via BQFE, the limitation of short-term stability by Dick effect is estimated as $2.4\times10^{-12}/\sqrt{\tau}$ according to the sensitivity function~\cite{6822984,710548} and the LO phase noises (See Appendix E).

To facilitate the comparison of two locking schemes without loss of generality, the dead time is ignored when calculating the Allan deviations.
The BQFE locking yields a stability of $4.3(2)\times10^{-12}/\sqrt{\tau}$, which is 5.1(4) dB better than the stability $1.4(1)\times10^{-11}/\sqrt{\tau}$ obtained by the PID locking.
By setting $\tau$ as unit time, the fractional stability $\sigma \propto \Delta f_{est}$.
Owing to the Heisenberg-scaling measurement of the clock frequency, the BQFE locking brings an enhancement of stability.
Our BQFE locking method could be extended to fountain clocks~\cite{Wynands2005fountain}, ion trap clocks~\cite{PhysRevLett.123.033201, PhysRevApplied.17.034041} and optical lattice clocks~\cite{Masao2005,Li_2024,Lu_2023,Luo_2020,Lin_2021}, which would be easier to implement than the cold-atom CPT clock as their transition probability can be directly measured via probing clock-state population.

\begin{figure}[ht]
	\includegraphics[width=\columnwidth]{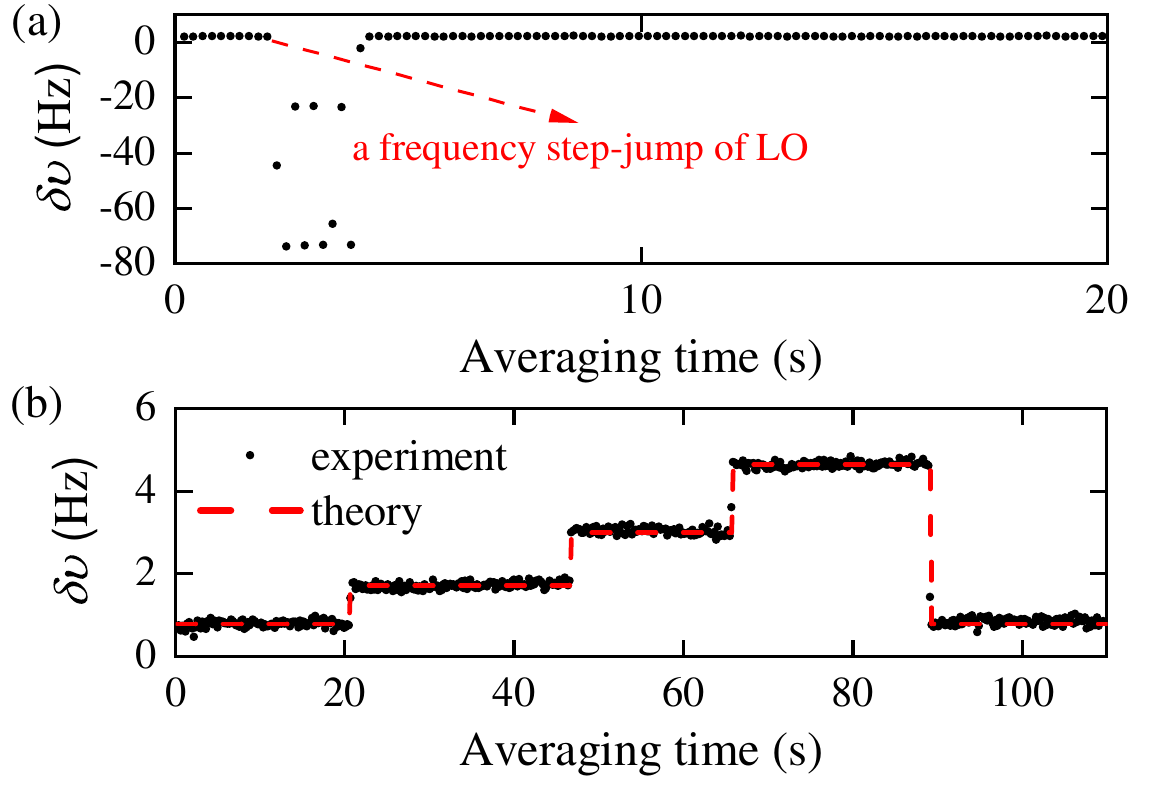}
\caption{\label{fig_track}
    (a) A time trace showing the closed-loop locking of a cold-atom CPT clock using BQFE when a frequency step change in the LO occurs suddenly. 
    This step-jump shifts the frequency outside the dynamic range of the Bayesian locking. 
    When the LO frequency gets back to normal, it afresh locks to the clock transition frequency.
    (b) A time trace showing the closed-loop locking of cold-atom CPT clock via BQFE when the bias magnetic field is changed. 
    The black dots are the experimental results for the field values of \{37, 55, 72, 90 and 37\} mG. 
    The red dash lines  correspond to the values calculated according to the second-order Zeeman coefficient.
    }
\end{figure}

In order to verify the locking capability of the BQFE, two different strategies are implemented in experiments.
(i) First, we verify the stability against a sudden frequency step-jump of LO that shifts the frequency outside the dynamic range of the Bayesian locking  [see Fig.~\ref{fig_track}~(a)].
When the clock is in normal operation, we fictitiously add a frequency offset into the LO.
As a result, the clock immediately loses the locking.
We keep the wrong LO frequency for 10 clock cycles, and the frequency of clock output is tangle-some.
When the LO frequency gets back to normal, the clock afresh locks to the clock transition frequency.
(ii) Second, we verify the trace ability of Bayesian locking to the clock transition frequency via changing the bias magnetic field [see Fig.~\ref{fig_track}~(b)].
When the clock is in normal operation, we suddenly change the bias magnetic field, therefore the clock transition frequency is accordingly changed due to the second-order Zeeman shift.
Our BQFE locking shows its robust trace ability when we step by step set the magnetic field as \{37, 55, 72, 90 and 37\} mG.
As expected, the experimental results are consistent with the theoretical values calculated according to the second-order Zeeman coefficient~\cite{Wu_2013}.

\begin{figure}[ht]
	\includegraphics[width=\columnwidth]{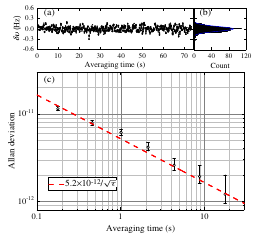}
\caption{\label{fig:feedback}
     Experimental results for the feedback after 9 Bayesian iterations (the total Bayesian iterations is 15).
    (a) The frequency fluctuation.
    (b) The corresponding histograms. 
    (c) The calculated Allan deviations.
    }
\end{figure}

The closed-loop locking of the cold-atom CPT clock is achieved by $13$ Bayesian iterations, no feedback is provided to the LO until completion of all $13$ Bayesian iterations. 
After $13$ Bayesian iterations and a feedback to the LO, we reset the BQFE procedure to implement the next feedback.
Therefore, the repetition time of clock operation is the total time of Bayesian iterations.
However, for high-stability atomic clock, the repetition rate is also important as the LO frequency may drift with time, especially the ultrastable laser locked on the ultra low expansion (ULE) cavity.
Here, we demonstrate that the repetition time of clock operation could be reduced by implementing the feedback in the Bayesian updates.
We experimentally verify the scenarios that the feedback is executed after $9$ Bayesian iterations when the BQFE procedure includes the total Bayesian iterations of 15.
This means that when the iteration number is larger than $9$, the $f_{est}$ is fed back to the LO at each subsequent iteration.
The experimental results are shown in Fig. \ref{fig:feedback}.
Although the calculated Allan deviations are just slightly deteriorated, there are a small number of data points that deviate the center of data distribution and the data distribution width is also broadening.
Hence, one should balance the repetition rate and the precision of the clock when implementing feedback in the Bayesian updates.

In our experiments, we set the total iterations as $13$, and reset the BQFE procedure after these iterations are done.
If we keep iterating and feeding back to the LO without reset operation, the distribution of the estimator will be ultra-narrow so that the new measurements have little effect on the estimator $f_{est}$.
In this case, despite the estimator $f_{est}$ is fed back to the LO, the atomic clock is pseudo-locked because the estimator $f_{est}$ is not the real clock frequency.
So a reset operation is essential.

\section{Conclusion and discussions}\label{Sec5}

In conclusion, we design an adaptive Bayesian quantum frequency estimation (BQFE) protocol to enhance the closed-loop locking of atomic clocks.
With the developed BQFE protocol, we experimentally demonstrate the Heisenberg-scaling measurement of the clock frequency with respect to the total interrogation time and achieve the closed-loop locking of a cold-atom CPT clock with better stability.
The BQFE is carried out to measure $^{87}{\rm Rb}$ clock frequency via exponentially step-wise increasing the interrogation time of CPT-Ramsey interferometry and adaptively adjusting the LO frequency during Bayesian updates.
In comparison to the conventional PID locking, our Bayesian locking shows an enhancement of 5.1(4) dB in the stability and better robustness against technical noises.
As our BQFE protocol operates based upon Ramsey interferometry, it can be easily extended to enhance the performance of versatile interferometry-based quantum sensors.

The precision in our experiment is limited by the coherence time $T_{max}$ of cold $^{87}\textrm{Rb}$ atomic gases.
Thus, the stability can be improved when long $T_{max}$ is available, such as, cold atoms in optical trap may have a $T_{max}$ up to tens of seconds~\cite{Probe_Xu,article_young}.
By using our BQFE, the stability of an Sr optical clock~\cite{article_young} may be improved from $5.2\times 10^{-17}/\sqrt{\tau}$ to $1.2 \times 10^{-17}/\sqrt{\tau}$.
Besides, to improve the efficiency, one can adopt Markov-chain Monte Carlo and particle filter method~\cite{Granade_2012, Wang2017,Qiu_2022} to efficaciously reduce the amount of calculations in Bayesian updates.
Furthermore, applying the BQFE to many-body quantum systems~\cite{Nolan2021,Pezze2018, qute202300329,PhysRevLett.111.090801,PhysRevX.11.041045,PhysRevResearch.6.023201}, the QPN can be further suppressed.
In particular, besides the cascaded GHZ-state phase estimation~\cite{Cao2024MultiqubitGA, Finkelstein2024}, the BQFE offers another efficient method to overcome the challenge of reduced dynamic ranges in entanglement-enhanced quantum sensors~\cite{Robinson2024DirectCO, article_Nichol,Edwin2020,pnas.0901550106,PhysRevLett.117.143004,PhysRevLett.102.033601,Bohnet2014,Esteve2008,Hosten2016MeasurementN1,Pezze2018,10.1063/5.0204102,Cao2024MultiqubitGA,Finkelstein2024}.

\acknowledgments{ C.H., Z.M. and Y.Q. contributed equally to this work. This work is supported by the National Key Research and Development Program of China (2022YFA1404104), and the National Natural Science Foundation of China (12025509, 12104521).}

\setcounter{equation}{0}
\setcounter{figure}{0}
\renewcommand{\theequation}{A\arabic{equation}}
\renewcommand{\thefigure}{A\arabic{figure}}

\section*{APPENDIX A: Basic procedure of our Bayesian locking}

Based upon the designed Bayesian quantum frequency estimation (BQFE) protocol, the atomic clock locking in our experiment is implemented according to the following flowing chart (see Algorithm \ref{BPE_algorithm}).
The basic procedure includes the following steps.
\begin{itemize}
\item Step 1: Determine the values of all input parameters $\{T_1,T_{max},R,a,g,\tilde{M},M_b, M_c\}$ based on specific experimental conditions and requirements.
\item Step 2: Initialize the frequency interval $[f_l, f_r]$ and prior distribution $p_1(f_c)$. 
The frequency interval should include the clock frequency $f_c$ (i.e. $f_l<f_c<f_r$) and the interval width equals to the reciprocal of the minimum integration time (i.e. $f_{lr} \equiv f_r-f_l=1/T_1$). Here, the start and end frequency is determined according to the estimated value as $f_l=f_{est}^{(i)}-f_{lr}/2$ and $f_r=f_{est}^{(i)}+f_{lr}/2$. 
The initial prior distribution is chosen as the uniform distribution over the interval $[f_l, f_r]$, i.e. $p_1(f_c)=1/T_1^{-1}$.
\item Step 3: Implement the Bayesian quantum frequency estimation loop: \\
(i) The Ramsey time $T_i$ is given by Eq.~(\ref{T_increase}). 
The interval width $f_{lr}$ is updated according to the reciprocal of the Ramsey time $T_i$, i.e. $f_{lr}=1/T_i$. 
For $i>1$, the frequency interval $[f_l, f_r]$ is updated by using the previous frequency estimator $f_{\rm est}^{(i)}$ as the center and $f_{lr}$ as the width, the prior distribution $p_i(f_c)$ is reset with the estimated center $f_{\rm est}^{(i)}$ and the estimated standard deviation $\Delta f_{\rm est}^{(i)}$ given by the previous step.\\
(ii) Obtain the frequency that maximizes Eq.~(5), defined as the LO frequency $f$.\\
(iii) Conduct experimental detection to obtain the normalized signal $s_f$ with $T_i$ and $f_i$ (i.e. the population probability $p_e$).\\
(iv) Perform Bayesian iteration. The likelihood function is defined by Eq.~(\ref{likelihood_ensemble}). 
The probability distribution is updated as a posterior distribution $p(f_c|p_e;f)$ according to Bayes’ formula Eq.~\eqref{Bayes_update}. 
The estimated value and uncertainty of $f_c$ can be obtained from the posterior distribution. 
The next update is implemented by inheriting the posterior distribution as the next prior distribution.\\
(v) After $M_b$ Bayesian iterations, output the estimated value $f_{est}^{j}$ and uncertainty $\Delta f_{est}^{j}$ of $f_c$
\item Step 4: Feedback to the LO. According to the the estimated value $f_{est}^{j}$, locking the LO frequency to the clock transition frequency via the error signal $f_{\rm est}^j-f_{\rm est}^{j-1}$.\\

Repetitive execution of Step 2 to Step 4 for the next feedback.
\end{itemize}

\begin{algorithm*}
\caption{Flowing chart of our Bayesian locking protocol}
\label{BPE_algorithm}
\SetKwInOut{Input}{Input}
\SetKwInOut{Output}{Output}
\SetKwInOut{Initialize}{Initialize}
\SetKwFunction{Actor}{Actor}
\SetKwFunction{Critic}{Critic}
\SetKwFunction{Softmax}{Softmax}
\SetKwFunction{CrossEntropy}{CrossEntropy}

\BlankLine
\Input{minimum interrogation time $T_1$;
maximum interrogation time $T_{\rm max}$;
other parameters $\{R, a, g, \tilde M, M_{b}\}$; total feedback number $M_c$.}
[Bayesian locking loop]:\\
\For{$j = 1$ \KwTo $M_{c}$}{
\Initialize{initial interval $[f_l, f_r]$; initial prior distribution $p_1(f_c) = 1/ T^{-1}_1$.}
\BlankLine
[Bayesian quantum frequency estimation Loop]:\\
\For{$i = 1$ \KwTo $M_{b}$}{
\BlankLine
[Updates of parameter]\;
interrogation time: $T_i=\left\{
    \begin{array}{ll}
	T_{\rm max}/ a^{(M_b-\tilde M-i)/g}, &  (M_b-\tilde M-i)/g \in \mathbb{N^+}, i < M_b-\tilde M\\
	T_{i-1}, &  (M_b-\tilde M-i)/g \notin \mathbb{N^+}, i < M_b-\tilde M\\
    T_{\rm max}, &  i \ge M_b-\tilde M
    \end{array} \right. $  \;
      length of interval: $f_{lr} = 1/T_i$\;
       \If{$i>1$}{
          $f_l \leftarrow f_{\rm est}^{(i)}-f_{lr}/2$\;
          $f_r \leftarrow f_{\rm est}^{(i)}+f_{lr}/2$\;
          reset the prior distribution $p_i(f_c) = \frac{1}{\sqrt{2\pi}\sigma} \exp{[- \frac{(f_c-\mu)^2}{2\sigma^2}] }$, where $\mu=(f_l+f_r)/2$ and $\sigma = \Delta f^{(i)}_{\rm est}$
       }
      LO frequency: $f = \arg{\max_{f}{U(f)}}$
      \BlankLine
      [Random enhancement]\;
      \If{$T_i>T_{\rm max}$}{$f \leftarrow f + \epsilon$, where $\epsilon \sim \mathcal{N}(0, 2\Delta f_c^{\rm est})$}
      \BlankLine
      [Experimental measurement]\;
      frequency shift compensation: $f \leftarrow f+f_s$\;
      measurement normalized signal $s_f$ using $T_i$ and $f_i$\;
      $p_e \leftarrow s_f$\;
      \BlankLine
      [Bayesian iteration]\;
      Likelihood function:    
        $\mathcal{L}(p_e\vert f_c, f)= \frac{1}{\sqrt{2\pi}\sigma}\exp{\left[-\frac{(p_e-\mathcal{L}_u(1\vert f_c,f))^2}{2\sigma^2}\right]}$,
      where $\sigma^2 = p_e(1-p_e)/R$\;
      Bayesian update: $p(f_c|p_e;f) \leftarrow \mathcal{N} \mathcal{L}(p_e|f_c, f) p(f_c)$\;
      Estimated frequency: $f_{\rm est} = \int{f_c p(f_c|p_e;f) \mathrm{d} f_c}$\;
      Uncertainty:
      $\Delta f_{\rm est} =
        \sqrt{\int{f_c^2 p(f_c|p_e;f) \mathrm{d} f_c} - (f_{\rm est})^2}$\;
    \Output{estimated frequency $f_{\rm est}^j$; uncertainty $\Delta f_{\rm est}^j$.}
    }
    Locking the LO frequency to the clock transition frequency via the error signal $f_{\rm est}^j-f_{\rm est}^{j-1}$.
    }
    \BlankLine
\end{algorithm*}

\begin{figure*}[ht]
    \includegraphics[width=\linewidth]{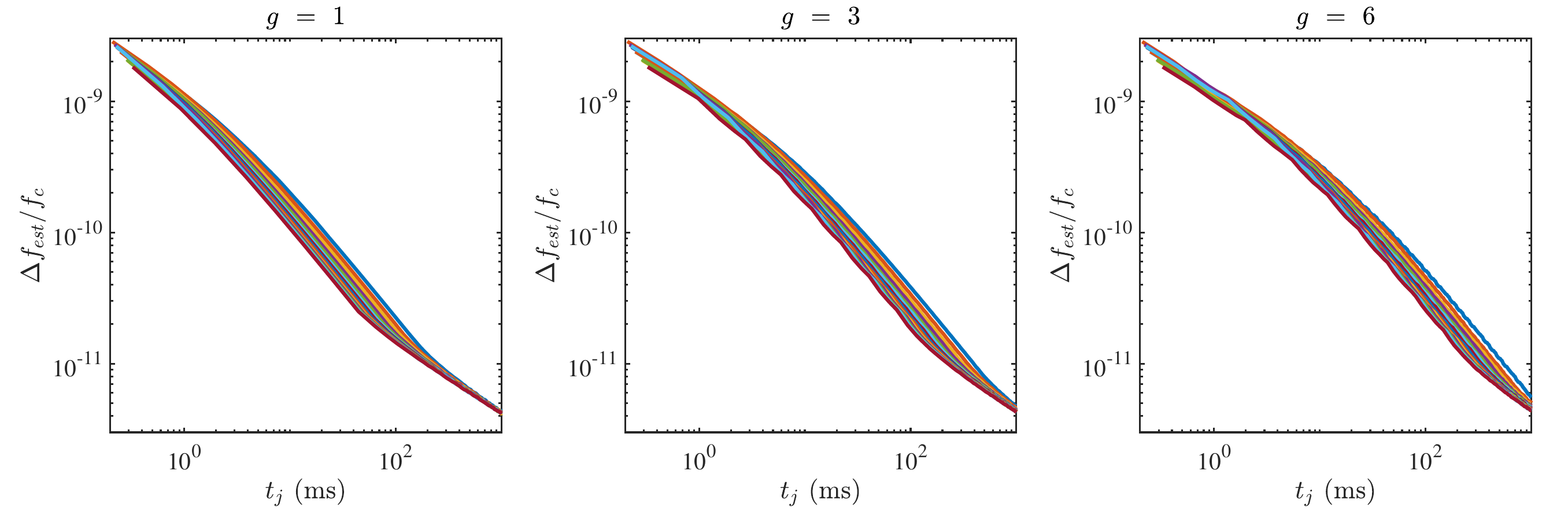}
    \caption{\label{sup_da}
  Simulated results of the estimation uncertainties $\Delta f_{est}$ with different $a$ for $g = 1$, $3$ and $6$.
    Different lines in each figure are results from different values of $a$ that is varied from 1.15 to 1.80 with a step of 0.05.
    }
\end{figure*}

\section*{APPENDIX B: Calculation of likelihood function and utilize function in our Bayesian quantum frequency estimation}

The uncertainty of the Gaussian function in the likelihood function Eq.~\eqref{likelihood_ensemble} is expressed as $\sigma^2  \approx p_e(1-p_e)/R = s_f(1-s_f)/R $ with $s_f$ the normalized Ramsey signal measured in the experiment and $R$ determined by the signal sensitivity.
Then we have
    \begin{equation}
        \frac{R_0}{R}=s_f = \frac{1}{2}\left[1+\cos({2\pi(f-f_c+f_s) T_R})\right],
    \end{equation}
    \begin{equation}
        \frac{R_1}{R}=1-s_f = \frac{1}{2}\left[1-\cos({2\pi(f-f_c+f_s) T_R})\right].
    \end{equation}
Thus,  
    \begin{equation}
        |\frac{\partial R_1}{\partial f_c}|= \pi R T_R \sin \left[{2\pi(f-f_c+f_s) T_R}\right],
    \end{equation}
For the binomial distribution, 
    \begin{equation}
        \Delta^2 R_1=R s_f (1-s_f) = \frac{R}{4} \sin^2 [{2\pi(f-f_c+f_s) T_R}].
    \end{equation}    
Therefore, we have 
    \begin{equation}\label{sensitivity}
        \Delta f_c= \frac{\Delta R_1}{|\frac{\partial R_1}{\partial f_c}|}=\frac{1}{ 2 \pi T_R \sqrt{R}}.
    \end{equation}      
According to Eq.~~\eqref{sensitivity}, we can determine $R=\frac{1}{(2\pi \Delta f_c T_R)^2}$ and the likelihood function can be calculated. 
The parameter $R$ is calculated according to Eq.~\eqref{sensitivity}, and the $\Delta f_c$ can be derived from the standard deviation of the frequency fluctuation of the conventional closed-loop locking, see the data in Fig.~\ref{fig_exp_clock}~(c). For our experiment, $R \approx 1540$.
 
Then, we use the prior distribution $p(f_c)$ to calculate the Utilize function $U_f$.
Here we give the details.
In the $(i-1)$-th step of BQFE algorithm we get a posterior distribution $p_{i-1}(f_c\vert p_e;f)$,
and we obtain a new prior distribution $p_{i}(f_c)=p_{i-1}(f_c\vert p_e;f)$,
whose Shannon information is:
	\begin{equation}
		U_0 = \int{p_{i}(f_c)\ln{[p_{i}(f_c)]}{\rm d}f_c}.
	\end{equation}
Then a posterior distribution $p_{i}(f_c\vert p_e;f)$ can be calculated using Bayes' formula:
	\begin{equation}
		p_i(f_c\vert p_e;f) \propto \mathcal{L}(p_e\vert f_c, f) p_{i}(f_c).
	\end{equation}
The gain of Shannon information is:
	\begin{equation}
		U_{p_e, f} = \int{p_{i}(f_c\vert p_e;f)\ln{[p_{i}(f_c\vert p_e;f)]}{\rm d}f_c} - U_0.
	\end{equation}
To get the expected gain in Shannon information,
we should integral $U_{p_e, f}$ over all possible values of $p_e$, which is from $0$ to $1$.
To perform the calculation in our experiment, we discretize the integral by dividing $L$ intervals,
then the likelihood function is~\cite{Dinani2019}:
    \begin{equation}\label{likelihood_r}
      \mathcal{L}(r|f_c, f)= \frac{1}{\sqrt{2\pi}\sigma}\exp[{-\frac{(r-L\mathcal{L}_u(1|f_c,f))^2}{2\sigma^2}}],
    \end{equation}
where $r=[p_e L]\in\{0,1,...,L\}$ and $\sigma^2=r(L-r)/R$.
Generally, it is appropriate to choose $L=R$ and $\sigma^2 \approx r(R-r)/R$.
However, if $R$ is large, the computation is time-consuming. 
In practice, we choose $L \ll R $ and this approximation is numerically valid when $L>20$.
    %It is also approximated from a binomial distribution when $R\gg 1$.
    %
The corresponding Bayesian update is thus:
    \begin{equation}
        p_i(f_c|r;f)=\mathcal{N} \mathcal{L}(r \vert f_c; f)p_{i-1}(f_c),
    \end{equation}
and the gain in Shannon information is:
\begin{equation}
        U_{r,f}=\int{p(f_c|r;f)\ln{[p(f_c|r;f)]}{\rm d}f_c} - \int{p(f_c)\ln{[p(f_c)]}{\rm d}f_c}.
\end{equation}    
    %  
%\begin{widetext}
In this way we can replace the integral by a summation:
    \begin{equation}
    \begin{split}
        &\int_{0}^{1}{\rm d}p_e{ U_{p_e, f} \int{p_{i}(f_c\vert p_e;f){\rm d}f_c}}\\
        &\propto
        \sum_{r=0}^L {U_{r,f}\int{p_{i}(f_c\vert r;f){\rm d}f_c}}
        \\
        &\propto
        \sum_{r=0}^L {U_{r,f} \int{\mathcal{L}(r\vert f_c, f)p_i(f_c){\rm d}f_c}}.
        \end{split}
    \end{equation}

We finally calculate the Utilize function $U_f$ by:
	\begin{equation}
			U_f = \sum_{r=0}^L {U_{r,f} \int{\mathcal{L}(r\vert f_c, f)p_i(f_c){\rm d}f_c}},
	\end{equation}
    %
    %where the value of $R$ is chosen according to the SNR.
    %
In our experiment we set $L=50$ as a trade-off between precision and computational cost.

\section*{APPENDIX C: Performance of our Bayesian quantum frequency estimation for different interrogation sequences}

\begin{figure*}[ht]
    \includegraphics[width=1.0\linewidth]{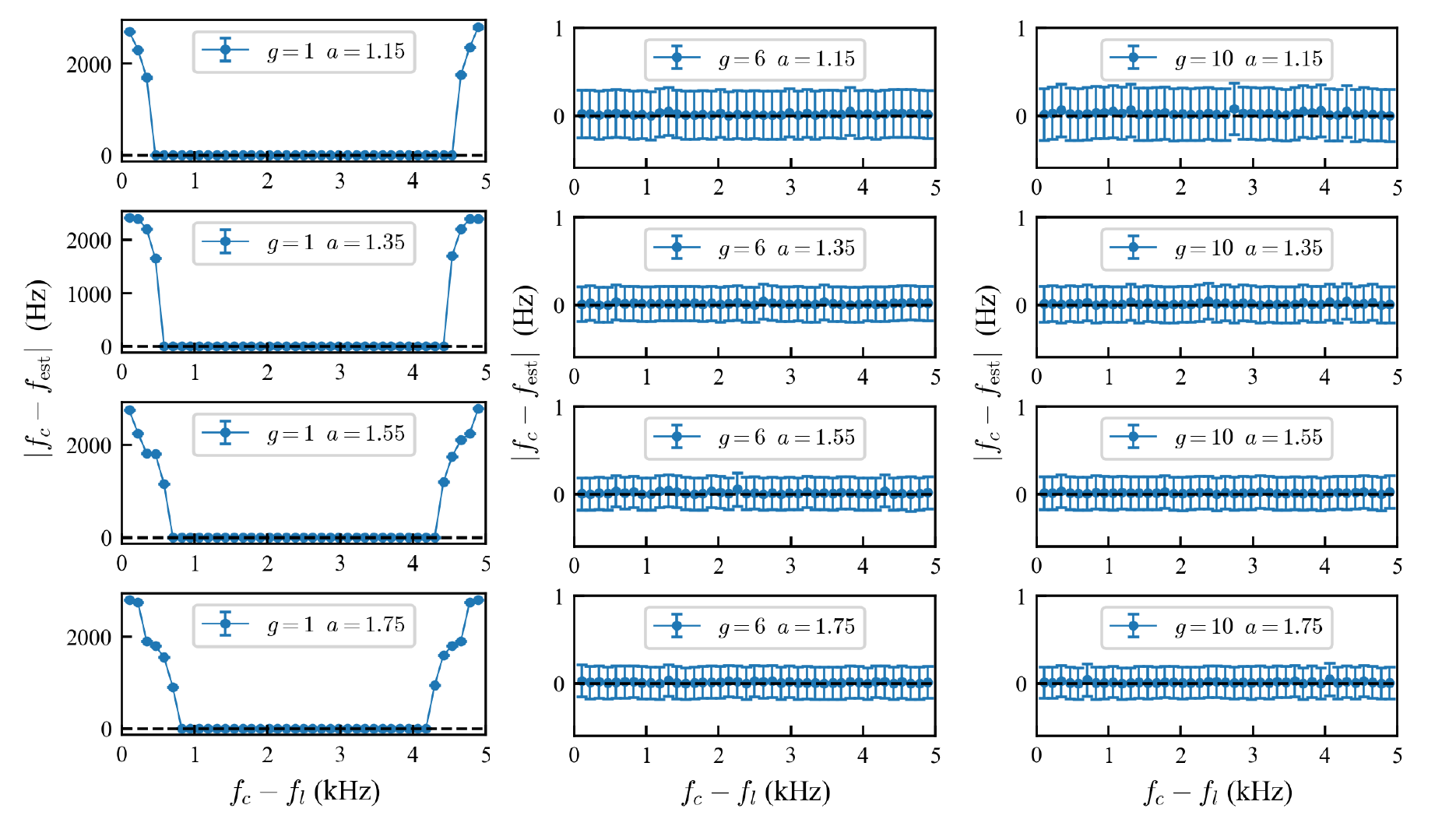}
    \caption{\label{dynamic}
    Simulated results of the absolute error $\vert f_c - f_{ est}\vert$ for $40$ different values when $g=10$.
    $g = 10$ means we change the value of $T_i$ once in every ten steps of Bayesian update and we set the value of $a$ as $1.15$, $1.35$, $1.55$ and $1.75$.
    }
\end{figure*}

Here, we consider a general increasing scheme for the interrogation time $T_i$ which is labelled by $\{a,g,\tilde M, M_b\}$. The $i$-th interrogation time can be expressed as
\begin{equation}\label{T_increase_A}
	T_i=\begin{cases}
	T_{\rm max}/ a^{\beta_i}, &  
 \beta_i \in \mathbb{N^+} \textrm{~and~} i < M_b-\tilde M,
 \\
	T_{i-1}, &  \beta_i \notin \mathbb{N^+} \textrm{~and~} i < M_b-\tilde M,\\
        T_{\rm max}, &  i \ge M_b-\tilde M,
	\end{cases}
\end{equation}
where the exponent $\beta_i=(M_b-\tilde M-i)/g$.

We show the influence of $a$ and $g$ in Eq.~\eqref{T_increase_A} and the simulated results using different $a$ and $g$ are shown in Fig.~\ref{sup_da}.
Here, we choose $g=1, 3, 6$.
Larger $a$ and smaller $g$ would enable a faster decrease of the standard deviations $\Delta f_{est}$ at the beginning.
For $g=1$, before $T_i$ reaches $T_{max}$, the standard deviations $\Delta f_{est}$ for larger $a$ is smaller as the scaling obeys the Heisenberg scaling $\Delta f_{est} \approx \frac{C}{T} \sqrt{1+2/(a-1)}$, as Eq.~\eqref{HL}. 
However, when $T_i$ attains $T_{max}$, the standard deviations $\Delta f_{est} \propto \frac{C}{\sqrt{T T_{max}}}$ for $\tilde M \gg \frac{a}{a-1}$.
This means that the standard deviations $\Delta f_{est}$ with different $a$ all converge to the same value at the end of iteration if the total interrogation time $T$ is sufficiently long.

Introducing the parameter $g$ so that the value of $T_i$ increases every $g$ steps  during Bayesian updates. 
In other words, $g=1$ means that the value of $T_i$ increases every step of Bayesian updates,
and $g=6, 10$ means the changing of $T_i$ happens in each $6$ or $10$ steps of Bayesian updates respectively.
We find that introducing $g$ can help to improve the dynamic range of using BQFE.

Since the minimum of $T_i$ is set as $0.2$ ms, the range of frequency $f_c$ to be estimated has a length of $5$ kHz.
In the numerical simulations, we select $40$ different values of $f_c$ uniformly-spaced in the $5$ kHz range,
and we use $f_l$ to represent the left end point of the range.
We use the same algorithm to estimate these $40$ values of $f_c$.
If the dynamic range is perfect, these $40$ estimators should have good consistency in terms of estimation error and uncertainty.
We choose $g=1,6,10$ and perform the numerical simulation as described above.
The simulation results are shown in Fig.~\ref{dynamic},
which indicates that the estimation performance for $g=1$ becomes worse when the value of $f_c$ is approaching the edges of the $5$ kHz range.
While for $g=6$ and $g=10$,
the degradation does not show up again, giving a good performance in the whole range of $5$ kHz.
It indicates that increasing the value of $g$ properly can improve the dynamic range of the frequency estimation.
It is also worth to mention that the left column in Fig.~\ref{dynamic} indicating the dynamic range can also be improved with smaller $a$ for the case of $g=1$. 
%\end{widetext}

\section*{APPENDIX D: Evaluation of the frequency shift induced by the probe light}

The electromagnetic field perturbation during probing induces frequency shifts on the atomic transitions known as light shifts.
The CPT clocks are known to experience significant frequency instabilities over medium to long time scales due to these light shifts~\cite{lin11lin_realization_3,10.1063/1.4977955,Limaojie_2024}.
Compared to the continuous interrogation scheme, the pulsed Ramsey interrogation scheme can reduce the light shifts.
However, the residual light shifts can still impact the accuracy of Ramsey spectroscopy.

\begin{figure}[ht]
    \includegraphics[width =\columnwidth]{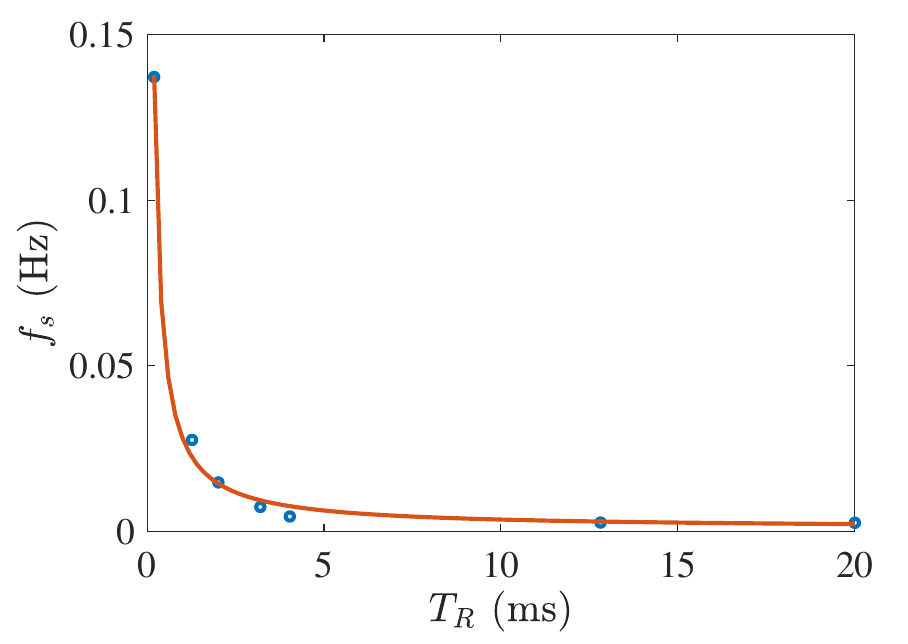}
    \caption{\label{frequency_shift}
    Dependence of the frequency shift $f_s$ on the interrogation time $T_R$.
    Dots are experimental data while the line is the fitting by $f_s =-\phi_{ls} / (2\pi T_R)$.
    }
\end{figure}

For the Ramsey interrogation, there are residual frequency shifts induced by the probe field during the interrogation pulses.
These interrogation-related shifts originate from a phase shift $\phi_{ls}$ accumulated during the interrogation pulses and inversely proportional to the Ramsey time as $f_s =-\phi_{ls} / (2\pi T_R)$.
In our BQFE protocol, the Ramsey time $T_R$ is required to increase in the Bayesian updates, which leads to different frequency of the atomic transition in the Bayesian iteration.
So the previous evaluation of the $T_R$-dependent frequency shifts $f_s$ are performed to guarantee the accuracy of estimator.
We measure these frequency shifts of different $T_R$ via locking the LO frequency to the central Ramsey fringe by a PID feedback.
The phase shift $\phi_{ls}$ is constant by keeping the preparation and detection pulses same for different interrogation time $T_R$.
The corresponding frequency shifts for different interrogation time $T_R$ are shown in Fig.~\ref{frequency_shift}.
Therefore, we can compensate these light shifts into the LO frequency, which can guarantee the accuracy of estimator.

\begin{figure}[ht]
    \includegraphics[width=\linewidth]{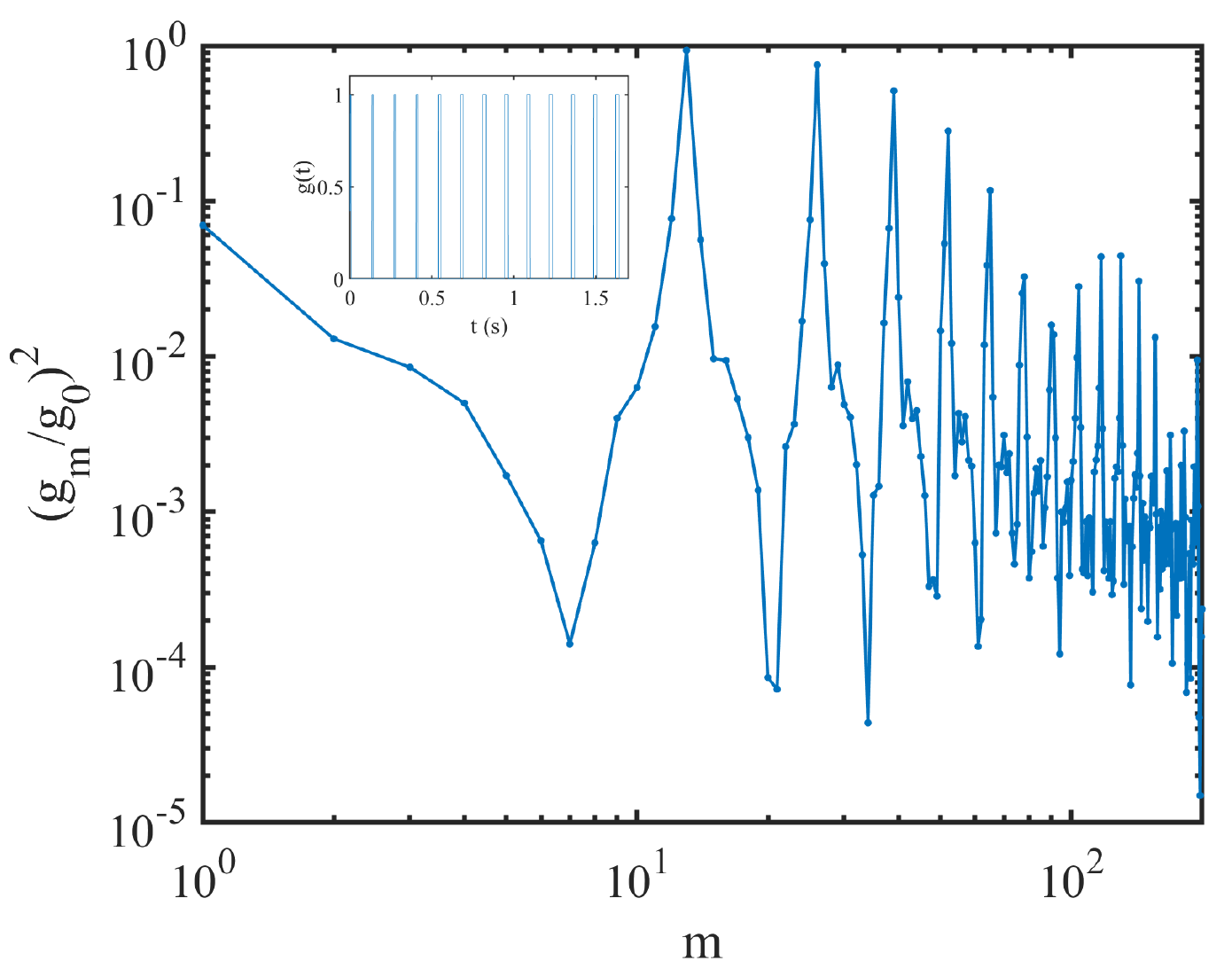}
    \caption{\label{dick_effect}
    The value of $(g_m/g_0)^2$ for BQFE locking.
    Inset: the sensitivity function of $g(t)$ in one clock cycle time.
    }
\end{figure}

\begin{figure*}[htp]
\includegraphics[width=\linewidth]{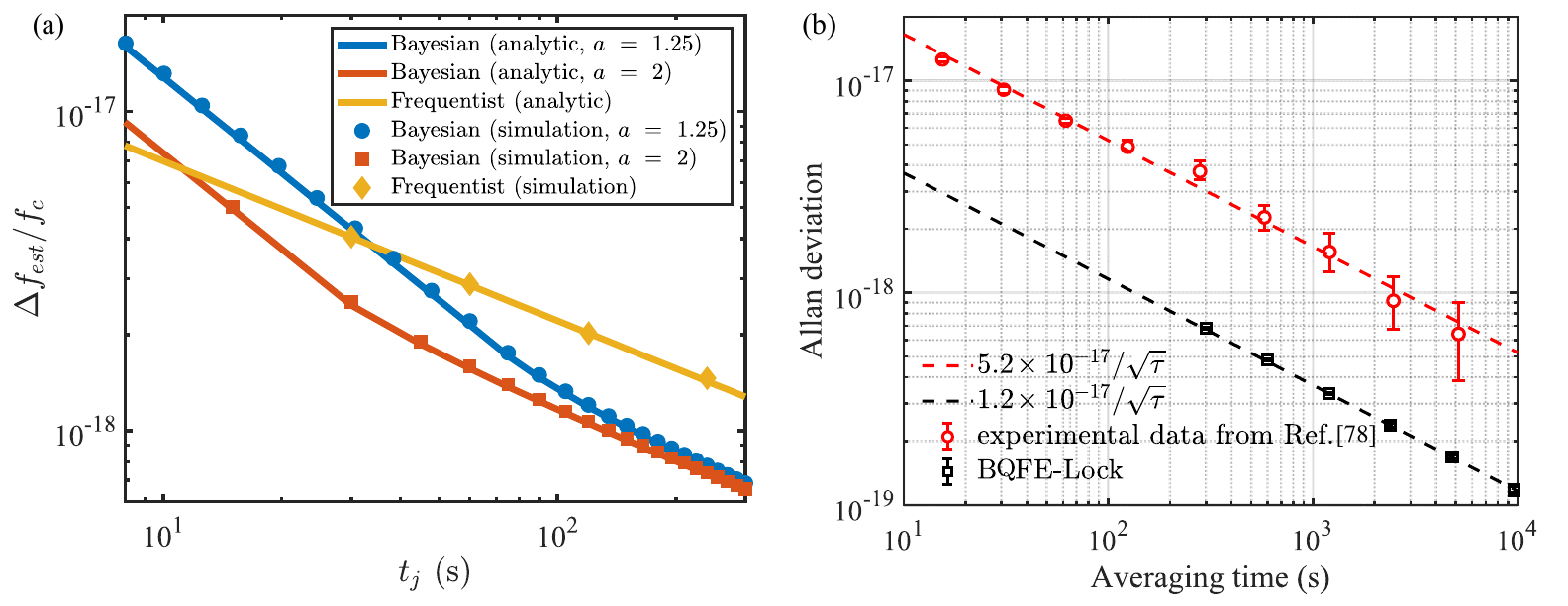}
\caption{\label{fig:Sr_f}
(a) The simulated precision via Bayesian quantum frequency estimation for Sr optical clock. The blue circles and orange squares are the simulated results using BQFE with $\{a=1.25,\ g=1,\ \tilde{M}=15,\ M_b=51\}$ and $\{a=2,\ g=1,\ \tilde{M}=18,\ M_b=30\}$, while the yellow rhombuses are obtained via frequentist measurement. The blue, orange, and yellow solid lines are the corresponding analytic results. 
(b) The simulated Allan deviation via Bayesian quantum frequency estimation for Sr optical clock. The black squares are the simulated results using BQFE with $\{a=1.25,\ g=1,\ \tilde{M}=15,\ M_b=51\}$, while the red circles are the stability of a Sr optical clock in Ref.~\cite{article_young}. The black and red dashed lines separately average down with slopes of $1.2\times10^{-17}/\sqrt{\tau}$ and $5.2\times10^{-17}/\sqrt{\tau}$. Here, $R=75$ and $T_{max}=15$ s.}
\end{figure*}

\section*{APPENDIX E: Dick effect}

For pulse-mode atomic clock, the LO frequency is periodically sampled by the atoms in the interrogation time as each clock cycle includes state preparation and detection (the dead time).
Therefore, the LO frequency noise at harmonic frequencies of the clock operation frequency is filtered by the atomic response and down-converted to low frequency noise, which is known as Dick effect and limits the clock stability~\cite{710548}.
The lowest limit to an achievable stability of the pulse-mode atomic clock induced by the LO frequency noise is given by,
	\begin{equation}\label{Dick effect limitation}
			\sigma^2(\tau) = \frac{1}{\tau}\sum^{\infty}_{m=1}[{\frac{g^2_m}{g^2_0}S^f(\frac{m}{T_c})]},
	\end{equation}
where $\sigma^2(\tau)$ is the fractional Allan variance for an averaging time $\tau$, $S^f(\frac{m}{T_c})$ is the one-sided power spectral density of the relative frequency fluctuations of the free-running LO frequency at Fourier frequency $m/T_c$.

Different from conventional Ramsey interferometry for two-level system, the sensitivity function $g(t)$ for CPT-Ramsey interferometry involving three atomic levels could be approximated as~\cite{6822984},
%\begin{widetext}
\begin{equation}\label{sensitivity function}
g(t)=\begin{cases}
	\exp[(t-\tau_p)(\gamma_C+1/\tau_O)],~~\textrm{if}~0\le t \le \tau_p, \\
	1,~~\textrm{if}~\tau_p\le t \le \tau_p+T_R+\tau_m, \\
     1-\frac{t-(\tau_p+T_R+\tau_m)}{\tau_d}, \\ ~~~~\textrm{if}~\tau_p+T_R+\tau_m \le t \le \tau_p+T_R+\tau_m+\tau_d, \\
     0, ~~\textrm{if}~\tau_p+T_R+\tau_m+\tau_d \le t \le T_B,
\end{cases}
\end{equation}
%\end{widetext}
where $T_B$ is the time of one iteration, $\tau_m$ is delay time before detecting the transmission signals, $\tau_O=\Gamma / \Omega^2 $ is the pumping time, $\gamma_C$ is the coherence relaxation rate.
For our BQFE locking, 13 Bayesian iterations are used then $T_c=13 T_B$.
The sensitivity function $g(t)$ is plotted in the inset of Fig.~\ref{dick_effect}.
$g_0$ and $g_m$ in Eq.~\eqref{Dick effect limitation} are defined from the sensitivity function $g(t)$ as
	\begin{equation}\label{g}
     \begin{split}
   & g_0=\frac{1}{T_c}\int_0^{T_c}g(t)dt,\\
   & g_m^2=(g_m^s)^2+(g_m^c)^2,\\
   & g_m^s=\frac{1}{T_c}\int_0^{T_c} \sin(2\pi mt/T_c)g(t)dt, \\
   & g_m^c=\frac{1}{T_c}\int_0^{T_c} \cos(2\pi mt/T_c)g(t)dt.\\
     \end{split}
	\end{equation}
And the value of $(g_m/g_0)^2$ for our BQFE locking is calculated and plotted in Fig.~\ref{dick_effect}.
The phase noise performances of the LO are -58 dBc/Hz, -68 dBc/Hz, -78 dBc/Hz and -90 dBc/Hz at 1 Hz, 10 Hz, 100 Hz and 1000 Hz, respectively.
According to Eq.~\eqref{Dick effect limitation}, the limitation of short-term stability by Dick effect is estimated as $2.4\times10^{-12}/\sqrt{\tau}$.

\section*{APPENDIX F: Feasibility of Bayes-enhanced strontium optical clock}

For an atomic clock with fixed averaging time $\tau$ and total interrogation time $T$, the fractional stability with Ramsey interferometry is proportional to the standard deviation, i.e., $\sigma(N,T,\tau) \propto \Delta f_{est}(N,T)$.
Therefore, benefit from the Heisenberg-limited sensitivity for $T$, the stability can be improved via BQFE locking. 
In our experiment, the precision limit is set by the limited coherence time $T_{max}=20$ ms of cold $^{87}\textrm{Rb}$ atomic gases under free falling.
If $T_{max}$ is longer, the stability can be further improved. 
For example, the cold atoms holding in optical trap may have a $T_{max}$ up to tens of seconds~\cite{Probe_Xu,article_young}.
Our BQFE locking can also be applied to the state-of-the-art optical clocks to improve the performances.

In Ref~\cite{article_young}, the tweezer-trapped $^{88}\textrm{Sr}$ had an optical-clock excited-state lifetime exceeding $40$ s in ensembles of approximately $150$ atoms.
By synchronous comparison between two sub-ensembles at the near-optimal Ramsey time of 15 s, they achieved the Allan deviation averaging down with a slope of $5.2(3) \times 10^{-17}/\sqrt{\tau}$.
To facilitate the comparison of the results reported in Ref~\cite{article_young} and our BQFE protocol, we use $T_{max}=15$ s and $R=75$ for simulation.
$T_{max}=15$ s corresponds to the near-optimal Ramsey time and $R=75$ is the atom number of sub-ensembles.

We use two different sets of parameters $\{a=1.25,\ g=1,\ \tilde{M}=15,\ M_b=51\}$ and $\{a=2,\ g=1,\ \tilde{M}=18,\ M_b=30\}$ to simulate the sensitivity $\Delta f_{est}$ of estimating the clock frequency, see Fig.~\ref{fig:Sr_f}~(a).
For larger $a$, the standard deviation $\Delta f_{est}$ decrease faster, but converge to the same value after several iterations when $T_i$ reaches $T_{max}$.
A sensitivity of $\Delta f_{est}=2.8 \times 10^{-4}$ Hz is achieved when total interrogation time $T=\sum_{i=1}^{M_b} T_i=300$ s.
Based on this point, a larger $a$ means a fewer iterations $M_b$ for the same sensitivity, which can reduce number of experimental measurements.
But one should make a trade-off as the estimated value $f_{est}$ is gradually converging to the real value.
By using the parameters $\{a=1.25,\ g=1,\ \tilde{M}=15,\ M_b=51\}$ for simulating the Bayesian locking of this tweezer clock, the stability may be improved from $5.2 \times 10^{-17}/\sqrt{\tau}$ to $1.2 \times 10^{-17}/\sqrt{\tau}$, see Fig.~\ref{fig:Sr_f}~(b), which provides experimental advantage and feasibility for strontium optical clock with BQFE locking.

%

%\newpage
%\newpage

\end{document}